\documentclass[aps,prd,nofootinbib,showkeys,reprint,floatfix,preprintnumbers]{revtex4-1}

\pdfoutput=1
\usepackage{graphicx}
\usepackage{graphics}
\usepackage[mathscr]{eucal}
\usepackage{amssymb}
\usepackage{amsmath}
\usepackage{xspace}
\usepackage{listings}
\usepackage{ulem}
\usepackage{color} 
\usepackage{graphicx}

\definecolor{brown}{rgb}{0.8,0.4,0}

\newcommand{\SARAH}[0]{{\tt SARAH}\xspace}
\newcommand{\SPheno}[0]{{\tt SPheno}\xspace}

\newcommand{\UBL}[0]{{\ensuremath{U(1)_{B-L}}}\xspace}

\newcommand{\vev}[0]{\textit{vev}\xspace}
\newcommand{\vevs}[0]{\textit{vev}s\xspace}

\newcommand{\EQ}[0]{Eq.\xspace}
\newcommand{\EQS}[0]{Eqs.\xspace}
\newcommand{\TAB}[0]{Table\xspace}
\newcommand{\FIG}[0]{Fig.\xspace}

\newcommand{\EG}{\textit{e.g}.\xspace}
\newcommand{\IE}{\textit{i.e}.\xspace}
\def\gsim{\raise0.3ex\hbox{$\;>$\kern-0.75em\raise-1.1ex\hbox{$\sim\;$}}}
\def\lsim{\raise0.3ex\hbox{$\;<$\kern-0.75em\raise-1.1ex\hbox{$\sim\;$}}}

\newcommand{\gy}{\ensuremath{g_{Y}}}
\newcommand{\gchi}{\ensuremath{g_{\chi}}}
\newcommand{\gychi}{\ensuremath{g_{Y \chi}}}
\newcommand{\mzp}{\ensuremath{M_{Z'}}}
\newcommand{\mz}{\ensuremath{M_{Z}}}

\newcommand{\sect}[1]{Sec.~\ref{#1}}

\definecolor{mkgreen}{rgb}{0.2,.70,.3}

\newcommand{\AddrBonn}{%
Bethe Center for Theoretical Physics \& Physikalisches Institut der
 Universit\"at Bonn, \\
53115 Bonn, Germany }

\newcommand{\AddrWur}{%
Institut f\"ur Theoretische Physik und Astronomie,
Universit\"at W\"urzburg\\
Am Hubland,
97074 W\"urzburg, Germany }

\begin{document}

\title{$SO(10)$ inspired gauge-mediated supersymmetry breaking}

\author{M.\ E.\ Krauss} \email{manuel.krauss@physik.uni-wuerzburg.de}

\author{W.\ Porod} \email{porod@physik.uni-wuerzburg.de}\affiliation{\AddrWur}

\author{F.\ Staub}\email{fnstaub@th.physik.uni-bonn.de}
\affiliation{\AddrBonn}

\keywords{supersymmetry; extended gauge symmetry; gauge mediation; LHC }

\pacs{12.60.Jv, 12.60.Cn, 14.80.Da, 14.70.Pw}

\preprint{Bonn-TH-2013-06}

\begin{abstract}
We consider a supersymmetric model motivated by a $SO(10)$ grand unified theory: 
the gauge sector near the supersymmetry scale consists of 
$SU(3)_c \times SU(2)_L \times U(1)_R \times U(1)_{B-L}$.  We embed 
this model in minimal gauge mediation and incorporate neutrino data via 
an inverse seesaw mechanism. Also in this restricted model, 
 the additional $D$ terms can raise the light Higgs mass
in a sizable way. Therefore, it is much easier to obtain  
$m_h \simeq 125$~GeV without the need to push the supersymmetry spectrum to 
extremely large values as it happens in  models with minimal supersymmetric standard model particle 
content only. We show that this model
predicts a diphoton rate of the Higgs  equal to or smaller than
the standard model expectation. We discuss briefly the collider phenomenology 
with a particular focus on the next to lightest supersymmetric particle in which
this model offers the sneutrino as an additional possiblity. Moreover, we point
out that, also in this model variant, supersymmetry can be discovered in $Z'$ decays 
even in scenarios in which the strongly interacting particles are
too heavy to be produced at a sizable rate at the LHC with 14 TeV.
In addition, we show that  lepton flavor violating 
observables constrain the size of the neutrino Yukawa couplings for which, in
particular, muon decays and $\mu-e$ conversion in heavy atoms are of
particular importance. Once these constraints are fulfilled the rates for
$\tau$ decays are predicted to be below the reach of near-future experiments.
\end{abstract}

\maketitle

\section{Introduction}
The LHC is rapidly extending our knowledge of
 the TeV scale. However, there is currently no hint of new physics 
beyond the standard model (SM), which leads to severe lower limits on the 
mass of new, especially colored, particles. 
One of the most popular model classes to extend the SM
 is supersymmetry (SUSY),
in particular, the minimal supersymmetric standard model (MSSM).
As the MSSM itself has over 100 free parameters, which are mainly 
part of the SUSY breaking sector, mechanisms of SUSY breaking, which depend only on a few parameters and which predict 
distinct relations among the different soft terms, have been studied. Those models trigger 
SUSY breaking in our visible sector by communicating with a hidden sector 
in which SUSY gets broken at the first place. Popular mechanisms of transmitting 
SUSY breaking from the hidden to the visible sector work either via gravity 
 like in supergravity \cite{Nilles:1983ge,Drees:2004jm}
 or via gauge interactions with so-called messenger fields like 
 in gauge-mediated supersymmetry breaking (GMSB) 
 \cite{Dine:1981gu,Dine:1981za,Dimopoulos:1981au,Nappi:1982hm,AlvarezGaume:1981wy,Dine:1993yw,Dine:1993qm,Dine:1994vc,Dine:1995ag}
 with six parameters. GMSB has the appealing feature that it is a flavor blind SUSY breaking. Hence, it
solves automatically the flavor problem if the SUSY breaking scale is not too high. 
In addition, the gravitino is the lightest supersymmetric particle (LSP) 
and usually the dark matter candidate in this kind of models. 
 
 However, both models are under big pressure 
 because of the observation of  a SM-like Higgs boson with mass of 125~GeV
\cite{ATLAS:2012ae,Chatrchyan:2012tx}. Even if this mass is below the absolute upper limit 
of about 132~GeV, which can be reached for the light Higgs mass in the most general MSSM \cite{Kant:2010tf}, 
it is already
very hard to explain it in the constrained MSSM with five parameters  \cite{Buchmueller:2011sw,Bechtle:2012zk,Buchmueller:2012hv} and demands
large SUSY breaking masses and especially a large mass splitting in the stop 
sector. This mass splitting is caused by large trilinear couplings. However, in GMSB, these terms 
are always small even if the $\mu/B_\mu$ problem of the GMSB is solved \cite{Komargodski:2008ax}.
That makes it even more difficult to obtain 
a Higgs mass in the correct range \cite{Draper:2011aa}. This caused increasing interest in nonminimal 
GMSB models, which involve also superpotential interactions between the matter and messenger sector
to create large  trilinear terms 
\cite{Shadmi:2011hs,Evans:2011bea,Evans:2012hg,Albaid:2012qk,Abdullah:2012tq,Craig:2012xp,%
Evans:2013kxa,Donkin:2012yn,Byakti:2013ti,Grajek:2013ola}.

Another possibility to reduce this tension between the Higgs mass and the simplest
constrained models is to extend the Higgs sector of the MSSM. The smallest possible extension is to
add a gauge singlet like in the next-to-minimal supersymmetric standard model (NMSSM)
(see Refs. \cite{Maniatis:2009re,Ellwanger:2009dp} and references therein). The singlet and the corresponding 
superpotential coupling to the Higgs doublets can significantly lift the upper limit 
on the light Higgs mass of $m_h < M_Z$ at tree level
in the MSSM by new $F$-term contributions \cite{Ellwanger:2006rm}.
It has been shown that, even in the  constrained NMSSM, a Higgs mass of 125~GeV 
can be explained \cite{Ellwanger:2011aa,Gunion:2012zd}. If one drops in addition the assumption of a $Z_3$
symmetry and considers instead the generalized NMSSM, these masses are obtained with even less
fine-tuning \cite{Ross:2012nr}. The same feature can also be observed in
models with Dirac instead of Majorana gauginos which usually come with an extended Higgs sector \cite{Benakli:2012cy}. 
There are also some hints
that the branching ratio of the observed particles do not 
completely agree with the SM expectations. Especially, the diphoton rate 
seems to be enhanced what is usually hard to explain in the context of the (constrained) MSSM \cite{Carena:2011aa,Benbrik:2012rm}. 
In contrast, this enhancement can much more easily be obtained in the NMSSM or 
its generalized version \cite{Ellwanger:2011aa,SchmidtHoberg:2012yy,King:2012tr,SchmidtHoberg:2012ip,Heng:2012at,Cao:2012fz}.

A second possibility is to consider models with extended gauge
structures which arise naturally in the
context of embedding the SM gauge group in a larger group such
as $SO(10)$ or $E_6$; see, \EG, Refs. 
 \cite{Cvetic:1983su,Cvetic:1997wu,Aulakh:1997ba,Aulakh:1997fq,
FileviezPerez:2008sx,Siringo:2003hh}. In those models, the upper bound on the
 lightest Higgs boson is also relaxed, this time due to additional $D$-term
 contributions
\cite{Haber:1986gz,Drees:1987tp,Cvetic:1997ky,Nie:2001ti,%
Ma:2011ea,Hirsch:2011hg}. Furthermore, they often provide the possibility to 
explain neutrino data: either via seesaw types I--III, which involve heavy states
\cite{Minkowski:1977sc,Yanagida:1979as,Mohapatra:1979ia,Schechter:1980gr}, or
via an inverse or linear seesaw with additional matter fields at the SUSY scale 
\cite{Bertolini:2009qj,DeRomeri:2012qd,Basso:2013jh}. 
Another interesting feature is that such 
models can also have a new gauge boson ($Z'$)  with a mass in the TeV range 
\cite{Malinsky:2005bi,DeRomeri:2011ie}. Hence, intensive searches for $Z'$ bosons
have been performed, and bounds on their mass have been set 
\cite{Aaltonen:2011gp,Abazov:2010ti,Collaboration:2011dca,%
Chatrchyan:2011wq}.  For reviews on various $Z'$ models,
see, \EG, Refs. \cite{Leike:1998wr,Langacker:2008yv}. 

In addition, $U(1)$ extensions of the SM provide another peculiar feature, which can 
have very interesting effects, namely, gauge kinetic mixing \cite{Holdom:1985ag,Babu:1997st,delAguila:1988jz}.
While gauge kinetic mixing is often ignored in phenomenological studies,
it has been shown recently in several works that it can have 
a significant effect on $Z'$ phenomenology \cite{Rizzo:1998ut,Rizzo:2012rf,Krauss:2012ku} 
but also on the Higgs mass \cite{O'Leary:2011yq} and dark matter properties 
\cite{Mambrini:2011dw,Basso:2012gz}.

In this work, we will assume minimal GMSB inspired by $SO(10)$: the grand unified theory (GUT) group gets broken to $SU(3)_c \times SU(2)_L \times U(1)_R \times U(1)_{B-L}$
very close to the GUT scale but well above the scale of the messenger fields that trigger GMSB. In contrast to previous 
studies, we assume, however, that the breaking down to the SM gauge groups takes place at the TeV scale, \IE, well below the 
lowest messenger scale. Hence, we study a messenger sector charged under $U(1)_R \times U(1)_{B-L}$, which will change our 
boundary conditions. Furthermore, we gain an enhancement 
of the mass of the light Higgs boson. 
In addition, we follow the setup of Ref. \cite{Hirsch:2012kv}, in which this model has been studied in gravity mediation, 
and assume additional gauge singlets present at the SUSY scale to incorporate neutrino masses and mixing via the inverse seesaw. 

The remainder of this paper is organized
as follows. In \sect{sec:model}, we briefly summarize the
main features of the model and its particle content. In 
\sect{sec:results}, we discuss the results for Higgs physics
under consideration, check the bounds coming from LFV observables on
the model parameters, and comment on the expected collider phenomenology. 
We conclude in \sect{sec:conclusions}.

\section{Aspects of the Model}
\label{sec:model}

\subsection{Particle content and superpotential}
\label{subsec:superpotential}
In this section, we discuss briefly the particle content and the
superpotential of the model under consideration. For a detailed discussion of the particle spectrum, we refer to Refs. \cite{Hirsch:2012kv, Hirsch:2011hg}.
The superpotential is given by
\begin{align}
\notag \mathcal W =& Y^{ij}_u{\hat{u}}^{c}_i\hat{Q}_j\hat{H}_u
- Y_d^{ij} {\hat{d}}^{c}_i\hat{Q}_j\hat{H}_d
- Y^{ij}_e {\hat{e}}^{c}_i\hat{L}_j\hat{H}_d
+\mu\,\hat{H}_u\hat{H}_d \\
& \label{eq:superpotentila}+Y^{ij}_{\nu}{\hat{\nu}}^{c}_i\hat{L}_j\hat{H}_u+Y_S^{ij} \hat \nu^c_i \hat S_j \hat \chi_R - 
\mu_R \hat{ \bar{ \chi}}_R \hat \chi_R +\mu_S^{ij}  \hat S_i \hat S_j\,,
\end{align}
where the upper line corresponds to the standard MSSM superpotential, and the lower line contains the new sector as well as the ingredients for the 
inverse seesaw mechanism: $Y_S $ and $Y_\nu$ being the neutrino Yukawa couplings and $\mu_S$ the mass term for the singlet field $S$, which is 
responsible for the mass of the light neutrinos. In \sect{subsec:tadpoles}, we will see
that this model with an inverse seesaw mechanism for neutrinos is much easier to implement in GMSB 
than the corresponding model with the type I seesaw in which the $\hat S_i$ fields are absent.
\begin{center}
\begin{table*}[ht]
\begin{tabular}{|c|c|c|c|c|c|c|} 
\hline \hline 
Superfield & Spin 0 & Spin \(\frac{1}{2}\) & Generations &
$SU(3)_c \times SU(2)_L$ &   $U(1)_R \times \UBL$ &   $U(1)_Y \times U(1)_{\chi}$\\ 
\hline \hline
\(\hat{Q}\) & \(\tilde{Q}\) & \(Q\) & 3
 & \(({\bf 3},{\bf 2}) \) & \((0,\frac{1}{6}) \) & \((\frac{1}{6},\frac{1}{4}) \) \\ 
\({\hat{d}}^{c}\) & \(\tilde{d}^c\) & \(d^c\) & 3
 & \(({\bf \overline{3}},{\bf 1})\) & \((\frac{1}{2},-\frac{1}{6}) \) & \((\frac{1}{3},-\frac{3}{4}) \)\\ 
\({\hat{u}}^{c}\) & \(\tilde{u}^c\) & \(u^c\) & 3
 & \(({\bf \overline{3}},{\bf 1})\) & \((-\frac{1}{2},-\frac{1}{6}) \) & \((-\frac{2}{3},\frac{1}{4}) \) \\ 
\(\hat{L}\) & \(\tilde{L}\) & \(L\) & 3
 & \(({\bf 1},{\bf 2})\) &\((0,-\frac{1}{2}) \) &\((-\frac{1}{2},-\frac{3}{4}) \)\\ 
\({\hat{e}}^{c}\) & \(\tilde{e}^c\) & \(e^c\) & 3
 & \(({\bf 1},{\bf 1})\) &\((\frac12,\frac{1}{2}) \) &\((1,\frac{1}{4}) \)\\ 
\({\hat{\nu}^{c}}\) & \(\tilde{\nu}^c\) & \(\nu^c\) & 3
 & \(({\bf 1},{\bf 1})\) &\((-\frac12,\frac{1}{2}) \) &\((0,\frac{5}{4}) \)\\ 
 \({\hat{S}}\) & \(\tilde{S}\) & \(S\) & 3
 & \(({\bf 1},{\bf 1})\) & \((0,0) \) & \((0,0) \)\\ 
\(\hat{H}_d\) & \(H_d\) & \(\tilde{H}_d\) & 1
 & \(({\bf 1},{\bf 2})\) &\((-\frac{1}{2},0) \) &\((-\frac{1}{2},\frac{1}{2}) \)\\ 
\(\hat{H}_u\) & \(H_u\) & \(\tilde{H}_u\) & 1
 & \(({\bf 1},{\bf 2})\) & \((\frac{1}{2},0) \) & \((\frac{1}{2},-\frac{1}{2}) \)\\ 
\(\hat{\chi}_R\) & \(\chi_R\) & \(\tilde{\chi}_R\) & 1
 & \(({\bf 1},{\bf 1})\) &\((\frac12,-\frac12) \) &\((0,-\frac{5}{4}) \)\\ 
\(\hat{\bar{\chi}}_R\) & \(\bar{\chi}_R\) & \(\tilde{\bar{\chi}}_R\) & 1
 & \(({\bf 1},{\bf 1})\) &\((-\frac12,\frac12) \) &\((0,\frac{5}{4}) \)\\ 
\hline \hline
\end{tabular} 
\caption{Chiral superfields and their quantum numbers with
respect to \( SU(3)_c\times\, SU(2)_L\times\,  U(1)_R\times\,
\UBL\). We also give the quantum numbers in the basis \( SU(3)_c\times\, SU(2)_L\times\,  U(1)_Y\times\,
U(1)_\chi\), the relations between both 
bases is defined in \sect{subsec:gkm}.}
\label{tab:matter_content}
\end{table*}
\end{center}
The scalar fields $\chi_R$ and $\bar \chi_R$ break  $U(1)_{R} \times U(1)_{B-L}$ 
down to $U(1)_Y$. 
As we interpret the $B-L$ charge of these fields as a lepton number, this leads to a spontaneous
breaking of the usual $R$ parity. Moreover, the usual $R$ parity would allow additional terms in the superpotential
such as $\hat{\bar{ \chi}}_R \,\hat{L}_j\,\hat{H}_u$, which also contribute to this
breaking as soon as electroweak symmetry is broken. To avoid this, we introduce
a $Z_2^M$ matter parity as it has also been proposed in Refs. \cite{Dev:2009aw,BhupalDev:2010he}
in similar frameworks. 
Under this parity, $\hat{H}_d$,
$\hat{H}_u$, $\hat{ \chi}_R$, and $\hat{\bar{ \chi}}_R$ are even, and all other fields are odd. 
We have checked that in this way also the contraints due to the so-called discrete gauge symmetry
anomalies are fulfilled \cite{Ibanez:1991hv, Dreiner:2005rd}. 
For completeness, we note that this symmetry is sufficient to forbid
the dangerous terms leading to proton decay which is the main purpose of the usual $R$ parity.
Moreover, also, the stability of the lightest supersymmetric particle is ensured in this way.

Interestingly, the particle content of this model
is in agreement with gauge coupling unification even if the breaking scale of
$SU(3)_c\times SU(2)_L \times U(1)_R \times U(1)_{B-L} \to 
SU(3)_c\times SU(2)_L \times U(1)_Y$ is close to the breaking scale down to
$SU(3)_c\times U(1)_{em}$. Therefore, we will always assume a one-step breaking
$SU(3)_c\times SU(2)_L \times U(1)_R \times U(1)_{B-L} \to SU(3)_c\times U(1)_{em}$ in the following. 
However, to facilitate the comparison with the MSSM, we will work in a different basis for the
$U(1)$ sector: we will take $U(1)_Y\times  U(1)_\chi$ as the orthogonal basis instead of
$U(1)_R \times U(1)_{B-L}$. We give the corresponding $U(1)$ quantum numbers for both bases
in \TAB~\ref{tab:matter_content}.

The soft SUSY breaking terms are 
\begin{align}
 \notag \mathscr V_{soft} &=  m^2_{ij} \phi_i^* \phi_j +\bigg( \frac{1}{2} M_{ab} \lambda_a \lambda_b
 + B_\mu H_u H_d + B_{\mu_R} \bar{{\chi}}_R \chi_R \\
 \notag &+B_{\mu_S} \tilde S \tilde S 
  + T_d^{ij} H_d \tilde d^c_i \tilde Q_j + T_u^{ij} H_u \tilde u^c_i \tilde Q_j  \\
&+ T_e^{ij} H_d \tilde e^c_i \tilde L_j
 + T_\nu^{ij} H_u \tilde \nu^c_i \tilde L_j + T_S^{ij} \chi_R \tilde \nu^c_i \tilde S_j 
 + \text{ h.c.} \bigg)\,,
\end{align}
with the generation indices $i$ and $j$. We have introduced here $\phi_i$ for all scalar 
particles and $\lambda_a$ for the different gauginos.
Note that, because of the two Abelian gauge groups present in the model
and the consequential gauge kinetic mixing discussed in the next subsection,
also the mixed soft gaugino term $M_{Y\chi} \lambda_{Y} \lambda_{\chi} $ is present
\cite{Fonseca:2011vn}. 

\subsection{Gauge kinetic mixing}
\label{subsec:gkm}
Even if $U(1)_{R}$ and $U(1)_{B-L}$ can be embedded orthogonal in $SO(10)$ 
at a given scale a kinetic mixing term of the form
\begin{align}
\label{eq:KM}
 \mathscr L_{mix} = -\chi F^{B-L\,,\mu \nu} F^R_{\mu \nu}
\end{align}
can occur. 
The reason is that the Higgs fields we 
assume to be present at the SUSY scale do not form a complete 
representation of $SO(10)$.
Hence kinetic mixing 
 will be introduced by renormalization group equation (RGE) evolution. 
This can be seen by the off-diagonal elements of the anomalous dimension matrix, which
in the basis $(U(1)_R, U(1)_{B-L})$ at one loop is given by
\begin{align}
 \gamma = \frac{1}{16 \pi^2} N \begin{pmatrix}
           \frac{15}{2} & \frac12 \\
           \frac12 & \frac{9}{2}
          \end{pmatrix} N\,.
\end{align}
Here, $N = \text{diag}(1,\sqrt{3/2})$ contains the GUT normalization.
In order to correctly account for gauge kinetic mixing effects, we follow the approach given in Ref. \cite{Fonseca:2011vn} and
shift the term to a covariant derivative of the form
\begin{align}
 D^{\mu} = \partial^\mu - i Q_{l}G_{lm}A_{m}^\mu \,,
\end{align}
where
\begin{align}
 G = \begin{pmatrix}
      g_R & g_{RBL} \\
      g_{BLR} & g_{BL}
     \end{pmatrix}\,,
\end{align}
 $A^\mu = (A^\mu_R, A^\mu_{B-L})^T$ and $Q$ is a vector containing the 
$U(1)$ charges of the field under consideration. 
We  assume the breaking into $U(1)_{R}\times U(1)_{B-L}$ to take place at the 
GUT scale $M_{GUT}$ and demand $g_{RBL}=g_{BLR}=0$ at $M_{GUT}$ as the initial condition.
In addition, we have the freedom to go into a particular basis by rotating 
the gauge bosons of the Abelian groups. As already mentioned, for an easier comparison
with the usual GMSB, we take the basis $U(1)_Y \times U(1)_\chi$ for which the first factor is the usual
hypercharge and the second one is the orthogonal one within $SO(10)$.
The gauge couplings and charges of 
$U(1)_R\times U(1)_{B-L}$ and $U(1)_Y\times U(1)_{\chi}$ are (without GUT normalization) related via 
\begin{align}
\notag  A^\mu \to A'^\mu &= \begin{pmatrix}
                       A^\mu_Y \\
                       A^\mu_{\chi}
                      \end{pmatrix}\,,
 ~~ Q \to Q' = \begin{pmatrix} 
 q_{B-L} + q_R \\
 \frac{3}{2} q_{B-L} - q_R
 \end{pmatrix}\,, \\
  G \to G' &= \begin{pmatrix}
      \gy & \gychi \\
      0 & \gchi
     \end{pmatrix}\,,
     \label{eq:basis_rotation}
\end{align}
with 
\begin{align}
 \notag &\gy = \frac{g_{BL} g_R - g_{BLR} g_{RBL}}{\sqrt{(g_{BLR} - g_R)^2 + (g_{BL}-g_{RBL})^2}} \,, \\
 \notag &\gchi = \frac{2}{5} \sqrt{(g_{BLR} - g_R)^2 + (g_{BL}-g_{RBL})^2} \,,\\
\notag &\gychi = \\ &\frac{2 (g_{BL}^2 + g_{BLR}^2) + g_{BLR} g_R +
g_{BL} g_{RBL} - 3 (g_R^2+g_{RBL}^2)}{5 \sqrt{(g_{BLR} - g_R)^2 + (g_{BL}-g_{RBL})^2}}\,. 
\label{eq:gauge_coupling_relations}
\end{align}
 
\subsection{GMSB boundary conditions}
\label{subsec:gmsb_boundaries}

In GMSB models it is assumed that supersymmetry breaking is generated by 
one or more superfields $\hat X_k$ living in a ``secluded'' sector.
We assume for simplicity that only one field $\hat X$ is present which is coupled
to a set of  messenger superfields $\hat{\Phi}_i$ via 
\begin{align}
 \mathcal W_{GM} = \lambda_i \hat X \,\hat \Phi_i \hat{\bar{\Phi}}_i\,.
\end{align}
Furthermore, it is assumed that the scalar and auxiliary components of $X$ receive a vacuum expectation value (\vev)
\begin{equation} \langle X \rangle = M + \theta^2 F \, ,
\end{equation}
and that it couples universally to $\hat{\Phi}_i$, implying that one
can set $\lambda_i=1$.
The supersymmetry breaking due to the
$F$-term \vev is communicated to 
the visible sector via the gauge interactions 
of the $\Phi_i$. Since we are interested in minimal gauge mediation without 
spoiling gauge coupling unification, we assume that the messenger fields 
form a complete $SO(10)$ multiplet,  \EG,  a {\bf 10}-plet.
This results in two $SU(2)_L$ doublets and two $SU(3)_c$ triplets below the $SO(10)$ scale
with suitable charges under the Abelian gauge groups, which are listed in \TAB \ref{tab:messengers}.
\begin{table}
\centering
 \begin{tabular}{|l|c|c|c|}
 \hline \hline
  & $SU(3)_c \times SU(2)_L$ & $U(1)_R\times \UBL$ & $U(1)_Y \times U(1)_\chi$ \\ \hline \hline
  $\hat \Phi_1$ & $({\bf 1},{\bf 2})$ & $(\frac12, 0)$ & $(\frac12, -\frac12)$\\
  $\hat{\bar{\Phi}}_1$ & $({\bf 1},{\bf 2})$ & $(-\frac12, 0)$ & $(-\frac12, \frac12)$\\
  $\hat \Phi_2$ & $({\bf 3 },{\bf 1 })$ & $(0,-\frac13)$ & $(-\frac13,-\frac12)$\\
  $\hat{\bar{\Phi}}_2$ & $({\bf \bar 3},{\bf 1 })$ & $(0,\frac13)$ & $(\frac13,\frac12)$\\ \hline \hline
 \end{tabular}
   \caption{Quantum numbers of the messenger fields in the respective bases.}
 \label{tab:messengers}
 
\end{table}

The SUSY breaking gaugino and scalar masses are generated via 1- and 2-loop diagrams, respectively
\cite{Giudice:1997ni,Giudice:1998bp,Martin:1996zb, ArkaniHamed:1998kj}.
Neglecting gauge kinetic mixing,
the boundary conditions for the SUSY breaking masses are given by \cite{Martin:1996zb}
\begin{align}
\label{eq:Boundary1}
 M_a =& \frac{g_a^2}{16 \pi^2} \Lambda \sum_i n_a(i) g(x_i)\,, \\
 \label{eq:Boundary2}
 m^2_k =& 2 \Lambda^2 \sum_a C_a(k) \frac{g_a^4}{(16 \pi^2)^2}\sum_i n_a(i) f(x_i)\,, 
\end{align}
with $\Lambda=F/M$, $x_i = |\Lambda/M|$,  and 
$g(x)$ and $f(x)$ are approximately 1 for $x \lsim 0.2$.
$g_a$ denotes the coupling of gauge group $a$, and $i$ runs over the messenger fields. $n_a(i)$ is the Dynkin 
index of the messenger with respect to the gauge group $i$. We use a normalization in which 
$n_a =1$ for the \textbf{10} of $SO(10)$. $C_a(k)$ is the quadratic Casimir invariant of the scalar field $k$. 

For a proper
treatment of gauge kinetic mixing, we use the substitution rules for Abelian 
groups given in Ref. \cite{Fonseca:2011vn}.
The resulting soft masses for the gauginos and scalars at the messenger scale 
 read 
\begin{align}
  \label{eq:BoundaryC1}
  M_{A \neq Abelian} =& \frac{g_A^2}{16 \pi^2} \Lambda \sum_i n_A(i) g(x_i)\,, \\
  M_{kl=Abelian} =&  \frac{1}{16 \pi^2} \Lambda \Big( \sum_i g(x_i) G^T N Q_i Q_i^T N G  \Big)_{kl}\,, \\
\notag   m_k^2 = \frac{2}{(16 \pi^2)^2}&  \Lambda^2   \Big(  \sum_{A\neq Abelian} C_A(k)g_A^4\sum_i f(x_i) n_A(i) \\
  &+ \sum_i f(x_i) (Q_k^T N G G^T N Q_i)^2 \Big)\,.
   \label{eq:gmsb_boundaries}
\end{align}
The trilinear soft SUSY breaking parameters are, as usual in minimal GMSB, 
essentially zero at the scale of gauge mediation. The singlet
$S$ is a special case because it is a gauge singlet and, thus, would have
a zero mass at this level. However, it gets a mass at the 3-loop level, which
can be estimated to be 
\begin{equation}
m^2_S \simeq \frac{Y_S^2}{16 \pi^2} \big( m^2_{\chi_R} + m^2_{\nu^c} \big)\,.
\label{eq:mS2}
\end{equation}
Obviously, this mass squared parameter is suppressed by an additional loop factor,
and RGE effects usually drive it to negative values at the electroweak scale. 
However, as can be seen in \sect{subsec:sneutrinoNLSP}, this is compensated by
an $F$ term proportional to $\mzp^2$ yielding a positive mass squared for the corresponding
mass eigenstates. 

For completeness, we note that one can explain the neutrino data by adjusting $\mu_S$ and taking
$Y_\nu$ as well as $Y_S$ diagonal. $\mu_S$ is a small parameter, which does not affect the collider
phenomenology. However, we will discuss in \sect{sec:LFV} the effect of nondiagonal
entries in $Y_\nu$ and $Y_S$ in the range compatible with neutrino data on rare lepton decays.

\subsection{Tadpole equations}
\label{subsec:tadpoles}
We decompose the neutral scalar fields responsible for gauge symmetry breaking as usual:
\begin{align}
 \notag &H_u = \frac{1}{\sqrt{2}}(\sigma_u + i \phi_u + v_u), \hspace{.34cm} H_d = \frac{1}{\sqrt{2}}(\sigma_d + i \phi_d + v_d), \\
 &\chi_R = \frac{1}{\sqrt{2}}(\sigma_R + i \phi_R + v_{\chi_R}), \, \bar{\chi}_R = \frac{1}{\sqrt{2}}(\bar \sigma_R + i \bar \phi_R + v_{\bar \chi_R}). 
\label{eq:higgs_states}
 \end{align}
We use the minimization conditions to determine the 
parameters $|\mu|^2,~|\mu_R|^2,~B_\mu$, and $B_{\mu_R}$:
\begin{align}
 \notag B_{\mu} =& \frac{t_\beta}{t_\beta^2 - 1} 
\Big(
 m_{H_d}^2 -m_{H_u}^2 + \frac{v^2}{4}c_{2 \beta} \big(g_L^2+\gy^2\\  &+(\gchi-\gychi)^2\big)
  + \frac{5 v_R^2}{8} c_{2 \beta_R} \gchi (\gchi-\gychi) \Big)
\,, \label{eq:tadpole_Bmu}\\
\notag B_{\mu_R} =&  \frac{t_{\beta_R}}{t_{\beta_R}^2-1}
\Big( m_{\bar \chi_R}^2 - m_{\chi_R}^2  
 -\frac{5 v^2}{8}c_{2 \beta} \gchi (\gchi-\gychi) \\ &+ 
 \frac{25 v_R^2}{16} c_{2 \beta_R} \gchi^2\Big)\,, \label{eq:tadpole_BmuR}\\
 \notag |\mu|^2 =&\frac{1}{ t_\beta^2 - 1}  \Big(
   m_{H_d}^2 -  m^2_{H_u} t_\beta^2 \\ \notag &- 
\frac{v^2}{8} \big(g_L^2+\gy^2 +(\gchi-\gychi)^2\big)(t_\beta^2-1)  \\&
+ \frac{5 v_R^2}{16} c_{2 \beta_R} (1+t_\beta^2) \gchi (\gchi-\gychi) 
\Big)\,, \label{eq:tadpole_Mu}\\ 
\notag |\mu_R|^2 =&\frac{1}{t^2_{\beta_R} - 1} \Big(
    m_{\bar \chi_R}^2 - m_{\chi_R}^2 t^2_{\beta_R} \\ \notag
    &+ \frac{5 v^2}{16}c_{2 \beta} (t^2_{\beta_R}+1)\gchi (\gchi-\gychi)  \\
    &- \frac{25 v_R^2}{32} (t^2_{\beta_R} - 1)\gchi^2
    \Big)\,,
 \label{eq:tadpole_MuR}
\end{align}
where $g_L$ is the $SU(2)_L$ gauge coupling, $t_x,\,c_x,\,s_x=\tan x,\,\cos x,\,\sin x$, whereas
$\tan \beta=\frac{v_u}{v_d},~\tan \beta_R = \frac{v_{\chi_R}}{v_{\bar \chi_R}},~v^2
=v_u^2 + v_d^2$, and $v_R^2 = v_{\chi_R}^2 + v_{\bar \chi_R}^2$. Note, that the corresponding 
terms can be generated by the Giudice-Masiero mechanism \cite{Giudice:1988yz}
and are thus free parameters in our context.

\begin{figure}[htbp]
\centering
 \includegraphics[width=\linewidth, angle=0]{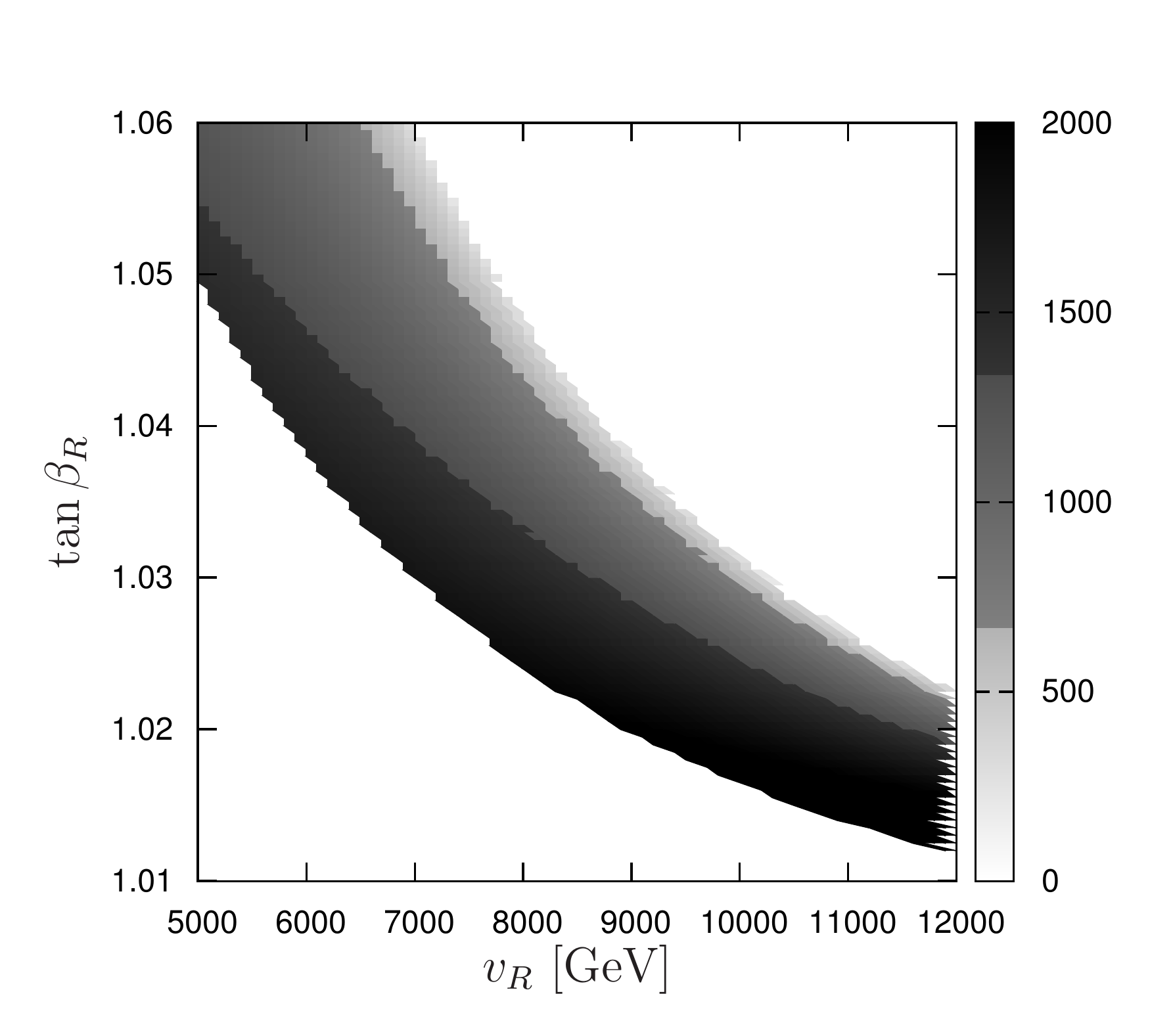}
 \caption{Allowed parameter space in the $v_R - \tan\beta_R$ plane. The plotted values correspond to $|\mu_R|$, which is calculated using the 
 tadpole equations. The free parameters have been set to $n=1,~\Lambda=5\cdot 10^5~\text{GeV}, ~M=10^{11}~\text{GeV}, ~\tan \beta=30,
 ~\text{sign}(\mu_R) = -,~diag(Y_S)
 =(0.7,0.6,0.6),~$ and $Y_\nu^{ii}=0.01$\,. 
 }
 \label{fig:vRtanBetaRplane}
\end{figure}

The latter of these equations is of particular interest as it is responsible for one of the 
major limitations to the model.
The 1-loop $\beta$ functions for the soft-breaking masses read in the limit of vanishing kinetic mixing
\begin{align}
\beta^{(1)}_{m^2_{\bar{\chi}}} =& -\frac{25}{2} \gchi^2 |M_\chi|^2 + \frac{5}{2} \gchi \sigma_\chi \label{eq:beta_chiRbar}\,,\\
\notag \beta^{(1)}_{m^2_{\chi}} =& -\frac{25}{2} \gchi^2 |M_\chi|^2 - \frac{5}{2} \gchi \sigma_\chi \\
&+ 2 \mbox{Tr}\Big( (m_{\chi}^2+m_{\nu}^2) {Y_S  Y_{S}^{\dagger}} + m_{S}^2  Y_{S}^{\dagger}  Y_S
     + T_{S}^*  T_{S}^{T}\Big)  \label{eq:beta_chiR}\,,
\end{align}
with 
\begin{align}
\notag \sigma_\chi =& \frac{\gchi^2}{4}\Big(5(m^2_{\bar{\chi}_R} -m^2_{\chi_R} ) + 4(m^2_{H_d}-m^2_{H_u})  \\
&+\mbox{Tr}\big(
m^2_{e^c} + 3 m^2_u + 5 m_{\nu^c}^2 + 6 (m_Q^2 - m_L^2) - 9 m_d^2
\big)
\Big)\,,
\end{align}
which is zero at the messenger scale and which stays zero if only 1-loop RGEs are used.
One can see that the main 
differences in the running are stemming from terms that are proportional to the trilinear soft-breaking 
couplings or the soft-breaking masses. 
Since we will consider the minimal GMSB where 
nonvanishing trilinear couplings are only generated via RGE evolution
and the breaking takes place well below the GUT scale, the splitting between the soft parameters 
$m_{\chi_R}^2$ and $m_{\bar \chi_R}^2$ will, in general, 
be smaller in comparison to a scenario with gravity mediation.  
Because of \EQ (\ref{eq:tadpole_MuR}), this immediately constrains
$\tan \beta_R$ to be larger than but close to one. The terms proportional to the \vevs squared then only give negative contributions to $|\mu_R|^2$,
\IE, there is an upper limit on  $|v_R|$ depending on $\tan \beta_R$ to find a solution to the tadpole equations. 
In Fig.~\ref{fig:vRtanBetaRplane} we show $|\mu_R|$ in the
$v_R$-$\tan \beta_R$ plane, in which one can see the correlation between the two
parameters. Note that, in the upper white area, one cannot achieve the correct gauge symmetry
breaking, whereas, in the lower white area, one encounters tachyonic states.

We have also considered the case where neutrino masses are generated by a seesaw type I mechanism 
similar to Ref. \cite{O'Leary:2011yq},
in which case there is no need to introduce the singlet field $\hat S$. 
Technically, this amounts in replacing
the terms $Y_S^{ij}\, \hat \nu^c_i\, \hat S_j\, \hat \chi_R\,  +\mu_S^{ij} \, \hat S_i\, \hat S_j$
in \EQ~(\ref{eq:superpotentila}) by $Y_S'^{ij} \hat{\nu}^c_i \hat{\chi}_R' \hat{\nu}^c_j$, where
$\hat{\chi}_R'$ has twice the $U(1)$ charges of $\hat \chi_R$. Performing the same chain of calculations,
one finds that there are hardly points with broken $U(1)_\chi$ as the larger gauge contributions in
the RGE evolutions
prevent $m_{\chi_R'}$ from becoming sufficiently small.

\subsection{Higgs sector}
\label{sec:modelhiggs}

In GMSB models with MSSM particle content, one needs a  SUSY spectrum in the
multi-TeV region to accommodate a Higgs mass of 125 GeV 
(see, \EG, Refs. \cite{Draper:2011aa, Ajaib:2012vc, Brummer:2012ns}). The reason is that the trilinear soft SUSY breaking
couplings are zero at the messenger scale in the minimal model, and thus the loop corrections to the
Higgs boson masses get reduced. In our model, the additional $U(1)$ factor gives already a sizable
$D$-term contribution to the tree-level part of the Higgs mass, and thus the need for large loop
corrections gets reduced.

On tree level, the scalar Higgs mass matrix in the basis $(\sigma_d, \sigma_u, \bar \sigma_R, \sigma_R)$ 
is given by 
\begin{widetext}
\begin{align}
\notag &m^2_{h^0} = \\
& \begin{pmatrix}
   \frac14 \tilde g_\Sigma^2 v^2 c_\beta^2 + m^2_A s_\beta^2
   & -\frac{s_{2 \beta}}{8}(\tilde g_\Sigma^2 v^2 + 4 m_A^2)
   & \frac{5}{8}\tilde g_\chi^2 v v_R c_\beta c_{\beta_R}
   & -\frac{5}{8}\tilde g_\chi^2 v v_R c_\beta s_{\beta_R}
   \\
   -\frac{s_{2 \beta}}{8}(\tilde g_\Sigma^2 v^2 + 4 m_A^2)
   & \frac14 \tilde g_\Sigma^2 v^2 s_\beta^2 + m^2_A c_\beta^2
   & -\frac{5}{8}\tilde g_\chi^2 v v_R s_\beta c_{\beta_R}
   & \frac{5}{8}\tilde g_\chi^2 v v_R s_\beta s_{\beta_R}
   \\
   \frac{5}{8}\tilde g_\chi^2 v v_R c_\beta c_{\beta_R}
   & -\frac{5}{8}\tilde g_\chi^2 v v_R s_\beta c_{\beta_R}
   & \frac{25}{16} \gchi^2 v_R^2c_{\beta_R}^2 + m_{A_R}^2s_{\beta_R}^2
   & -\frac{s_{2 \beta_R}}{32}(25 \gchi^2 v_R^2 + 16 m_{A_R}^2)
   \\
   -\frac{5}{8}\tilde g_\chi^2 v v_R c_\beta s_{\beta_R}
   & \frac{5}{8}\tilde g_\chi^2 v v_R s_\beta s_{\beta_R}
   & -\frac{s_{2 \beta_R}}{32}(25 \gchi^2 v_R^2 + 16 m_{A_R}^2)
   & \frac{25}{16} \gchi^2 v_R^2s_{\beta_R}^2 + m_{A_R}^2c_{\beta_R}^2
  \end{pmatrix}\,,
\label{eq:higgs_massmatrix}
\end{align}
\end{widetext}
where
${\tilde g_\Sigma^2} = g_L^2 + \gy^2 + (\gchi-\gychi)^2$,
$\tilde g_\chi^2 = \gchi (\gchi-\gychi)$, and $s_x,~c_x = \sin x,~\cos x$.
The parameters $m_{A}$ and $m_{A_R}$ are the tree-level masses of the pseudoscalar Higgs bosons, 
which are given by
$m^2_{A} = B_\mu / (s_{\beta} c_{\beta})$ and $m^2_{A_R} = B_{\mu_R} / (s_{\beta_R} c_{\beta_R})$. 
\begin{figure}[htbp]
\begin{minipage}{\linewidth}
 \includegraphics[width=\linewidth]{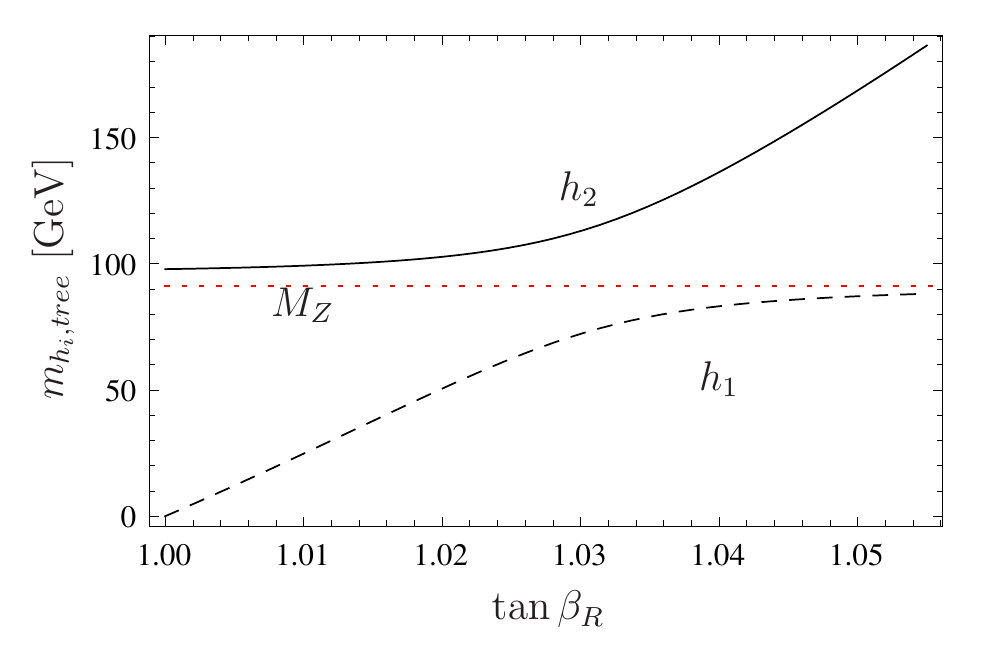}
 \end{minipage}
 \begin{minipage}{\linewidth}
 \includegraphics[width=\linewidth]{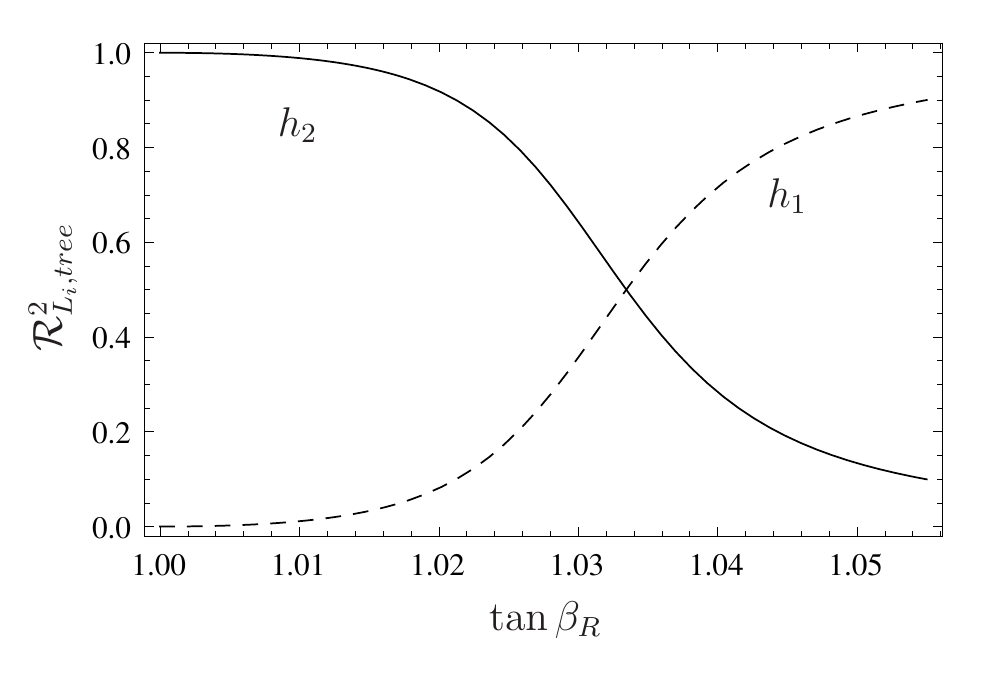} 
 \end{minipage}
 \caption{Tree-level dependence of the lightest Higgs masses (above) as well as the admixture of the $SU(2)_L$ doublet Higgses 
 $\mathcal R^2_{L_i}=|U_{i1}|^2+|U_{i2}|^2$ (below) on $\tan \beta_R$ 
 with the parameter choice of \FIG \ref{fig:vRtanBetaRplane} and $v_R = 7$~TeV.
The horizontal small dashed (red) line shows the $Z$ mass.}
 \label{fig:higgs_treelevel}
\end{figure}
Already at tree level, this leads to non-negligible contributions to the doublet Higgs mass
as $\frac14 \tilde g_\Sigma^2 v^2 \simeq M^2_Z + \frac14 v^2 (\gchi-\gychi)^2 > M^2_Z$. 
As typical values, we find $\gchi-\gychi \simeq 0.27$. This immediately
gives an upper bound on the tree-level Higgs mass:
\begin{equation}
m_{h,tree} \le \mz^2 + \frac14  (\gchi - \gychi)^2 v^2\,.
\label{eq:higgs_upper_bound}
\end{equation}
In Fig.~\ref{fig:higgs_treelevel}, we show the 
dependence of the two lightest Higgs states on 
$\tan \beta_R$ at tree level. 
Even in this very restricted model, the tree-level mass can easily reach 100 GeV while at the
same time requiring that this state is mainly an $SU(2)_L$ doublet Higgs boson. Even though the details
are changed by loop corrections, see, \EG, Ref. \cite{Hirsch:2011hg}, this figure also shows that 
$\tan\beta_R$ has to be close to 1 to obtain this desired feature. In the numerical part,
we include the complete 1-loop correction to \EQ~(\ref{eq:higgs_massmatrix}) and the
dominant 2-loop corrections to the MSSM sub-block.

\subsection{Dark matter}

As already mentioned, the gravitino is the LSP in GMSB models, and all SUSY particles decay into it in a cosmologically short time 
\cite{Dimopoulos:1996gy,Fujii:2002yx,Jedamzik:2005ir}.
The abundance of thermally produced gravitinos is under assumptions
consistent with the standard thermal evolution of the early Universe given by 
\begin{equation}
\label{eq:relic}
\Omega_{3/2} h^2 = \frac{m_{3/2}}{\mbox{keV}} \frac{100}{g_\star} \, \, .
\end{equation} 
Here, \(g_\star\) is the effective number of degrees of freedom at the time of gravitino decoupling. 
For a mass of $O(100)$~eV, the gravitino would form warm dark matter and would have the correct abundance to explain the 
observed dark matter relic density in the Universe. However, there are stringent
constraints on the contribution of warm dark matter from observations of the Lyman-\(\alpha\) forest \cite{Viel:2005qj}. 
These bounds rule out pure warm dark matter scenarios with particle masses below 8 keV for nonresonantly produced dark matter
\cite{Boyarsky:2008xj}. If one takes this lower limit into account, one sees that gravitinos, which have once been in thermal 
equilibrium, would overclose the Universe. This is known as the cosmological gravitino problem. 
There have been some proposals in literature to circumvent this problem by, for instance, additional entropy production after the 
freeze-out of the gravitino \cite{Baltz:2001rq,Fujii:2002fv,Lemoine:2005hu}. 
However, it turned out that entropy production from messenger decays hardly works \cite{Staub:2009ww}. 
Hence, one has to assume either other mechanisms like saxion decays \cite{Hasenkamp:2010if} or decays of moduli fields \cite{Kamada:2011ec}. 
Also, if the gravitino mass is in the MeV range, they might never have been in thermal equilibrium if the reheating temperature 
is sufficiently low \cite{Choi:1999xm}. Because of these 
very model dependent
issues, we do not address the question of the gravitino relic density in the following.

\section{Numerical results}
\label{sec:results}
\begin{center}
\begin{table*}[htbp]
 \begin{tabular}{|l|c|c|c|c|c|c|}
 \hline \hline
  & BLRI & BLRII & BLRIII & BLRIV & BLRV & BLRVI\\ \hline \hline
  $n$ & \multicolumn{3}{|c|}{4} & 1 & 1 & 1\\
  $\Lambda$ [GeV] & \multicolumn{3}{|c|}{$2.5\cdot 10^5$} & $5\cdot10^5$ & $3.8\cdot 10^{5}$ & $5 \cdot 10^{5}$\\ 
  $M$ [GeV] & \multicolumn{3}{|c|}{$10^{11}$} & $10^{10}$ & $9 \cdot 10^{11}$ & $10^{11}$\\ 
  $\tan \beta$ & \multicolumn{3}{|c|}{$40$} & 30 & 30 & 20\\ 
  $\tan \beta_R$ & \multicolumn{3}{|c|}{$1.04$} & 1.03 & 1.05 & 1.02\\ 
  sign$(\mu_R)$ & \multicolumn{3}{|c|}{$-$} & $+$ & $-$ & $+$\\ 
  $v_R$ [TeV] & \multicolumn{3}{|c|}{$7$} & 7.5 & 6.7 & 12\\ 
  $Y_\nu^{ii}$ & \multicolumn{3}{|c|}{$0.01$} & 0.01 & 0.01 & 0.01\\
  $diag(Y_S)$ & (0.65,0.65,0.1) & (0.65,0.65,0.3)  & (0.65,0.65,0.65) & (0.6,0.6,0.6) & (0.77,0.73,0.45) & (0.7,0.6,0.6)  \\ 
  \hline
  $m_{h_1}$ [GeV] & 70 & 92 & 125 & 70 & 108 & 98\\
  $\mathcal R^2_{L,h_1}$ & 0.006 & 0.018 & 0.961 & 0.003 & 0.094 & 0.006\\
  $m_{h_2}$ [GeV] & 126 & 127 & 156 & 124 & 124 & 124\\
  $\mathcal R^2_{L,h_2}$ & 0.994 & 0.982 & 0.039 & 0.997 & 0.906 & 0.995\\
  \hline
  $\mzp$ [TeV]& \multicolumn{3}{|c|}{$2.53$} & 2.7 & 2.41 & 4.32\\
  \hline  
  $m_{\nu_{h,1}}$ [GeV] & 357 & 1070 & 2306 & 2277 & 1542 & 3633\\
  $m_{\nu_{h,2}}$ [GeV] & 2309 & 2308 & 2306 & 2278 & 2497 & 3633\\
  $m_{\nu_{h,3}}$ [GeV] & 2309 & 2308 & 2306 & 2278 & 2633 & 4238\\
  \hline
  $m_{\tilde \nu_1}$ [GeV] & 334 & 909 & 1715 & 1728 & 1207 & 1863\\
  $m_{\tilde \nu_2}$ [GeV] & 1072 & 1546 & 1715 & 1757 & 1482 & 1879\\
  $m_{\tilde \nu_3}$ [GeV] & 2090 & 2048 & 1715 & 1759 & 1514 & 1879\\  
  \hline
  $m_{\tilde \tau_1}$ [GeV] & 906 & 906 & 905 & 867 & 764 & 1007\\
  $m_{\tilde \mu_R}$ [GeV] & 1166 & 1166 & 1165 & 976 & 877 & 1061\\
  $m_{\tilde e_R}$ [GeV] & 1167 & 1166 & 1166 & 976 & 877 & 1061\\  
  \hline
  $m_{\tilde \chi^0_1}$ [GeV] & 505 & 766 & 1156 & 575 & 453 & 589\\
  $m_{\tilde \chi^0_2}$ [GeV] & 1157 & 1157 & 1353 & 610 & 825 & 1043\\
  \hline
  $m_{\tilde \chi^\pm_1}$ [GeV] & 2216 & 2216 & 2217 & 1113 & 883 & 1142\\
  $m_{\tilde \chi^\pm_2}$ [GeV] & 2591 & 2590 & 2588 & 1956 & 1600 & 2015\\  
  \hline
  $m_{\tilde g}$ [GeV] & 5460 & 5459 & 5456 & 3018 & 2423 & 3076\\
  $m_{\tilde t_1}$ [GeV] & 4209 & 4209 & 4206 & 2993 & 2231 & 2941\\ 
  \hline \hline
 \end{tabular}
 \caption{Input parameters and mass spectrum of different representative parameter points.
}
 \label{tab:parameter_points}
\end{table*}
\end{center}
\subsection{Implementation in \SARAH and \SPheno}
We used the implementation of the model in \SARAH 
\cite{Staub:2008uz,Staub:2009bi,Staub:2010jh,Staub:2012pb} and 
\SPheno \cite{Porod:2003um,Porod:2011nf} presented in Ref. \cite{Hirsch:2012kv}
and extended it by the GMSB boundary conditions: here, we
allow for up to four messenger {\bf 10}-plets with degenerated masses.
At the messenger scale, we 
implemented the GMSB boundary conditions for the soft masses
using the \EQS~(\ref{eq:BoundaryC1})--(\ref{eq:gmsb_boundaries}).
The link between \SARAH and \SPheno 
allows for a precise mass spectrum calculation based on full 2-loop RGE running and the 
1-loop corrections to all masses. In addition, the known 2-loop corrections to the 
Higgs masses in the MSSM are linked \cite{Dedes:2003km,Brignole:2002bz,Brignole:2001jy,Degrassi:2001yf}. For more details of the mass spectrum 
calculation as well as the inclusion of SUSY thresholds, we refer to Ref. \cite{Hirsch:2012kv}. 
In addition, the \SPheno version created by \SARAH includes also routines for a full 1-loop 
calculation of the LFV observables $l_i\to l_j \gamma$, $l_i \to 3 l_j$, $\mu - e$ conversion in atoms,
flavor violating $\tau$ decays to a lepton and meson, and
$B_s\to \mu^+\mu^-$ \cite{Dreiner:2012dh}. 

For further discussions, we choose six benchmark scenarios BLRI--BLRVI which provide 
distinct features. These benchmark points are given in \TAB~\ref{tab:parameter_points} and 
will be discussed in the following subsections. 


\subsection{Higgs physics}
\begin{center}
\begin{figure*}[htbp]
\begin{minipage}{.49\linewidth}
 \includegraphics[width=\linewidth]{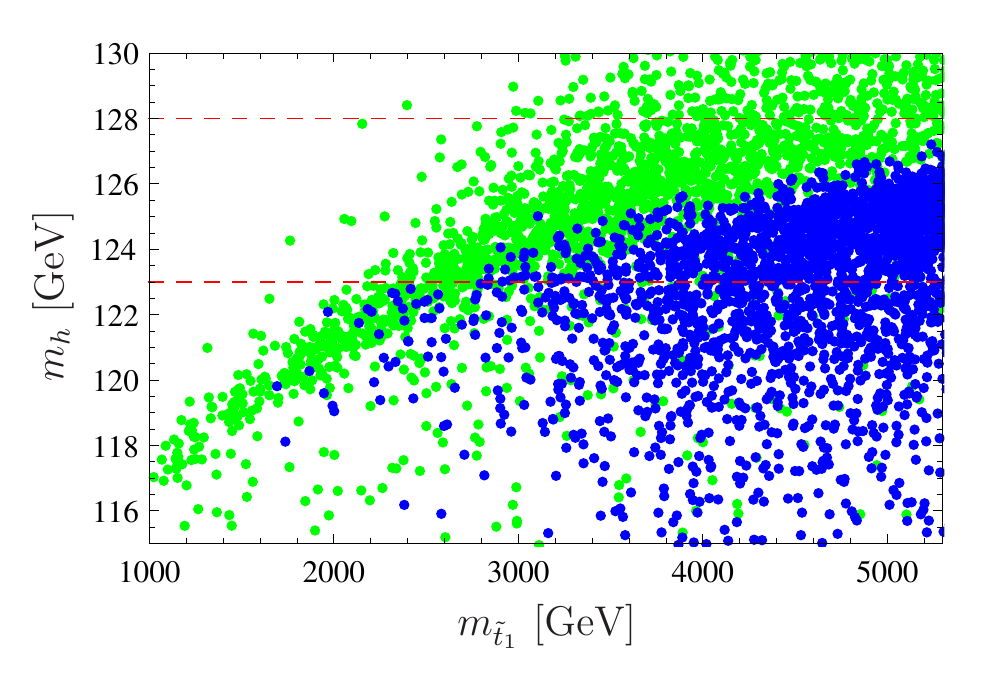}
 \end{minipage}
 \begin{minipage}{.49\linewidth}
 \includegraphics[width=\linewidth]{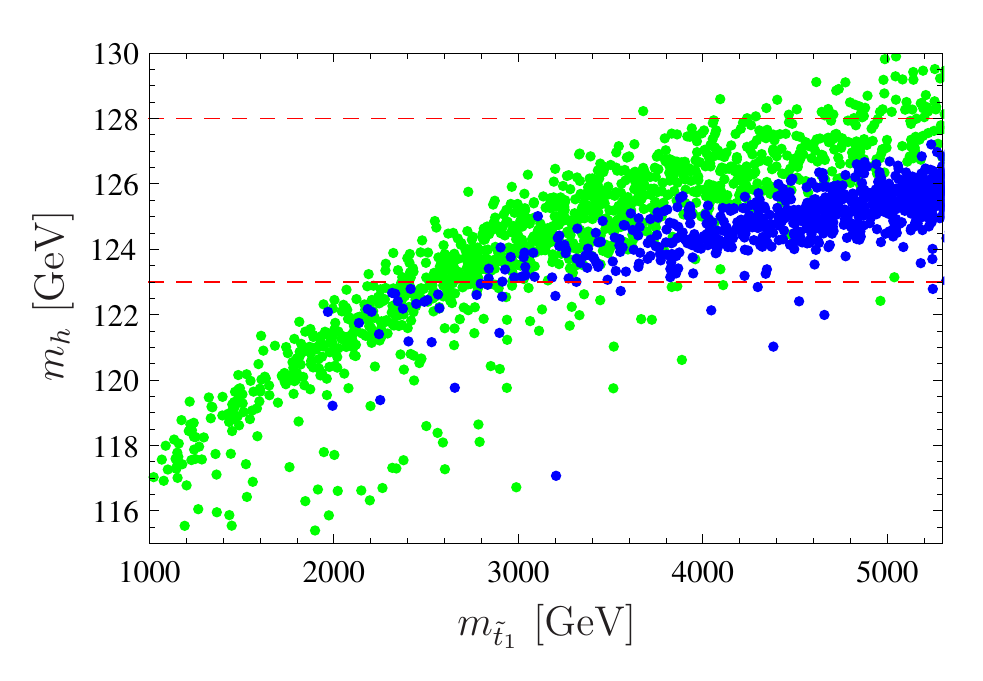}
 \end{minipage}
 \caption{Mass of the doublet-like Higgs versus
  the mass of the lightest
 stop for $n=1$ and the other
 parameters as in \TAB \ref{tab:scatter_scan}. 
 Only points with $R_{h \to \gamma \gamma}>0.5$ (left) and 0.9 (right) were included.  
 The blue dots represent points where the lightest eigenstate
 is doublet-like, green dots where it is the second-lightest Higgs. 
 }
 \label{fig:higgs_stop_scatter}
\end{figure*}
\end{center}

\begin{table}[hbt]
 \begin{tabular}{|c | c |}
  \hline \hline
  Parameter & Varied range \\ \hline \hline
  \# Messenger multiplets $n$ & 1~...~4 \\ 
  Messenger scale  $M$ & $(10^5~...~10^{12})$ GeV \\
  $\Lambda = F/M$ & $\frac{1}{\sqrt n} (10^5~...~10^6)$ GeV \\ 
  $\tan \beta$ & 1.5~...~40 \\ 
  $\tan \beta_R$ & 1~...~1.15 \\
  sign$(\mu_R)$ & $\pm 1$ \\
  $v_R$ & $(6.5~...~10)$ TeV \\
  $Y_S^{ii}$ & 0.01~...~0.8 \\
  $Y_\nu^{ii}$ & $10^{-5}$~...~0.5 \\ \hline \hline
 \end{tabular}
 \caption{Parameter ranges of the scan. The sign of $\mu$ has always been taken positive.}
 \label{tab:scatter_scan}
\end{table}
We performed a scan over the free parameter space in order to numerically check how well Higgs data can be 
accommodated for in our GMSB framework. The parameter variations can be found in \TAB \ref{tab:scatter_scan}. 
In \FIG \ref{fig:higgs_stop_scatter} we show the masses of the doubletlike Higgs vs the mass of the
lightest stop.
As expected from the discussion in \sect{sec:modelhiggs}, 
points where $h_2$ is the doubletlike Higgs are of 
particular interest since they allow for higher 
values of the Higgs mass at a fixed $m_{\tilde t}$. Due to the tree-level
contributions from the new sector, we can achieve the 
observed Higgs mass even for stop masses of about 2~TeV while a 
doubletlike $h_1$ requires $m_{\tilde t_1} \gtrsim 3~$TeV. 
Admittedly, this is quite a high scale in terms of naturalness in SUSY. However, compared to the lower limit of 
$m_{\tilde t} \gtrsim 5~$TeV in usual GMSB scenarios
(see, \EG, Ref. \cite{Ajaib:2012vc}) this is significantly lower. 
Such a heavy stop will be difficult to study at the LHC and will
potentially require a center of mass (c.m.)\ energy larger than 14 TeV. However,
here, an $e^+ e^-$ collider like CLIC with up to 5 TeV c.m.\ energy
might be an ideal machine to discover and study such a heavy stop; see,
e.g., Refs. \cite{Accomando:2004sz,Linssen:2012hp} and references therein. 

\begin{figure}[htbp]
\centering
\begin{minipage}{\linewidth}
 \includegraphics[width=\linewidth]{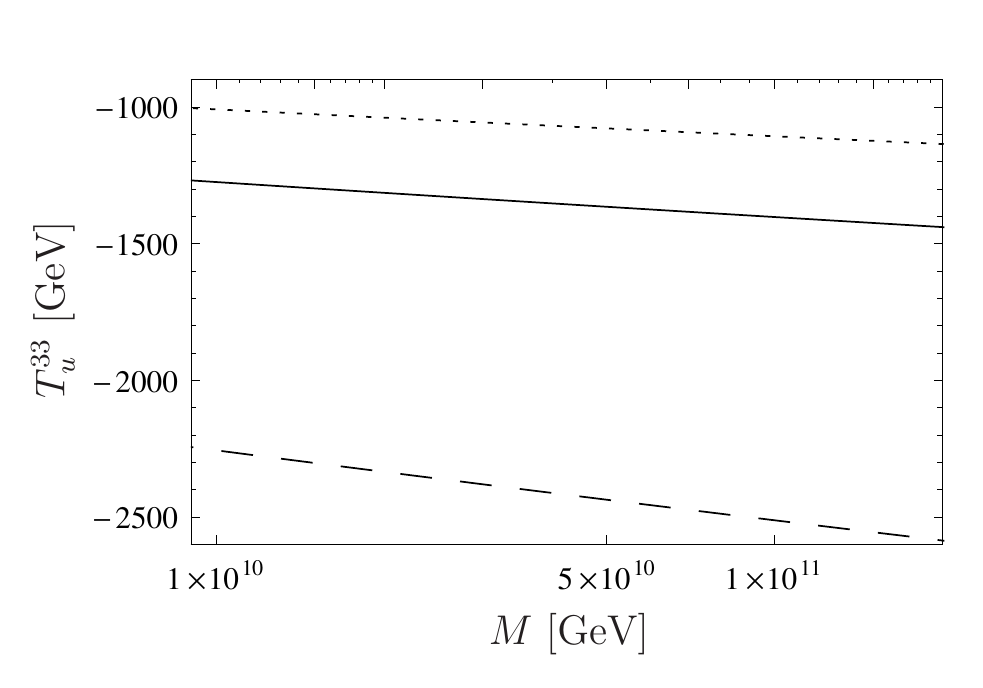} 
 \end{minipage}
 \begin{minipage}{\linewidth}
 \includegraphics[width=\linewidth]{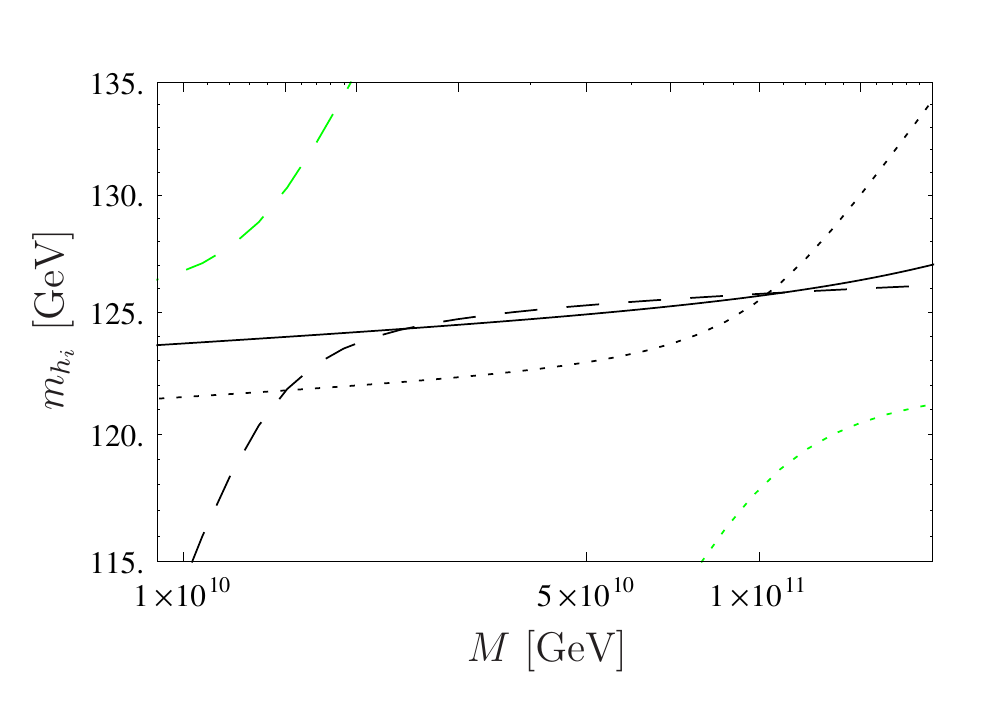}
 \end{minipage}
 \caption{$T_u^{33}$ and mass of the 
 doubletlike Higgs vs $M$
 for  BLRIII (black dashed line), BLRIV (black full line), and BLRV (black dotted
 line) (but for $\tan \beta_R=1.03$).
 The light green lines 
 correspond to the $\chi_R$-like Higgs state of the 
 corresponding parameter point.
}
 \label{fig:higgs_scale}
\end{figure}

A way to allow for a lighter SUSY spectrum in GMSB scenarios apart from 
the mixing with the extended Higgs sector is going up
to higher messenger scales, thus allowing a longer RGE running and hence larger
induced $T$ parameters as demonstrated in \FIG \ref{fig:higgs_scale}.
At $M\simeq 10^{10}$ and $10^{11}$~GeV there is a level crossing
between the two light states for the points BLRIII and BLRV, respectively,
which is the reason for the observed increase of $m_{h}$.

An interesting observable is the
rate $h\to \gamma\gamma$ as there are some hints for an enhancement above
SM expectations \cite{Aad:2012tfa,Chatrchyan:2012ufa,Hubaut}. 
We define the ratio $R_{h \to \gamma \gamma}$ by
\begin{align}
 R_{h \to \gamma \gamma} = 
 \frac{[\sigma (pp \to h)\times BR(h\to \gamma \gamma)]_{BLR}}{[\sigma (pp \to h)\times BR(h\to \gamma \gamma)]_{SM}}\,.
\end{align}
The cross sections for the main production channels, gluon fusion and vector boson fusion,
are essentially the SM-production cross section 
reweighted by the (effective) couplings of the Higgs boson $c_{hXX}^{BLR}$ normalized 
to the SM expectations $c_{hXX}^{SM}$:

\begin{align}
\notag \sigma (XX \to h)_{BLR} =&  \sigma (XX \to h)_{SM} \left(\frac{c_{hXX}^{BLR}}{c_{hXX}^{SM}}\right)^2\,, 
 \\ X=&g,W\,.
\end{align}
The main contribution to Higgs production comes from gluon fusion. The effective Higgs coupling to two gluons is completely determined
in the SM by the top and $W$ loop.

\begin{figure}[tbp]
\centering
  \includegraphics[width=\linewidth]{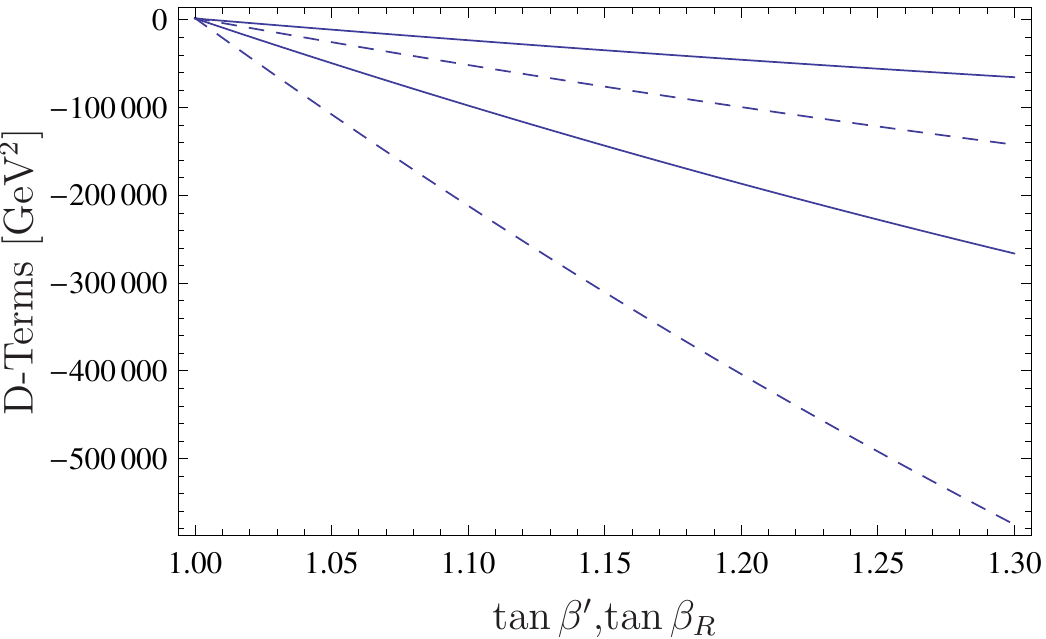}
 \caption{$D$ term contribution to the mass entries of the $R$ sleptons for
 $\tan\beta=10$, $\mzp=1.5,3.0$~TeV, and fixing the gauge couplings by the
 requirement of gauge coupling unification: $g^{Y\times BL}_{BL} = 0.55$, $g_Y = 0.36$, respectively, $g^{R\times B-L}_{BL} = 0.57 $, $g_R = 0.45 $.
The full (dashed) lines correspond to the $U(1)_R\times U(1)_{B-L}$ ($U(1)_Y\times U(1)_{B-L}$)
scenario.
}
 \label{fig:Dterms}
\end{figure}

\begin{figure}[htbp]
\centering
 \includegraphics[width=\linewidth]{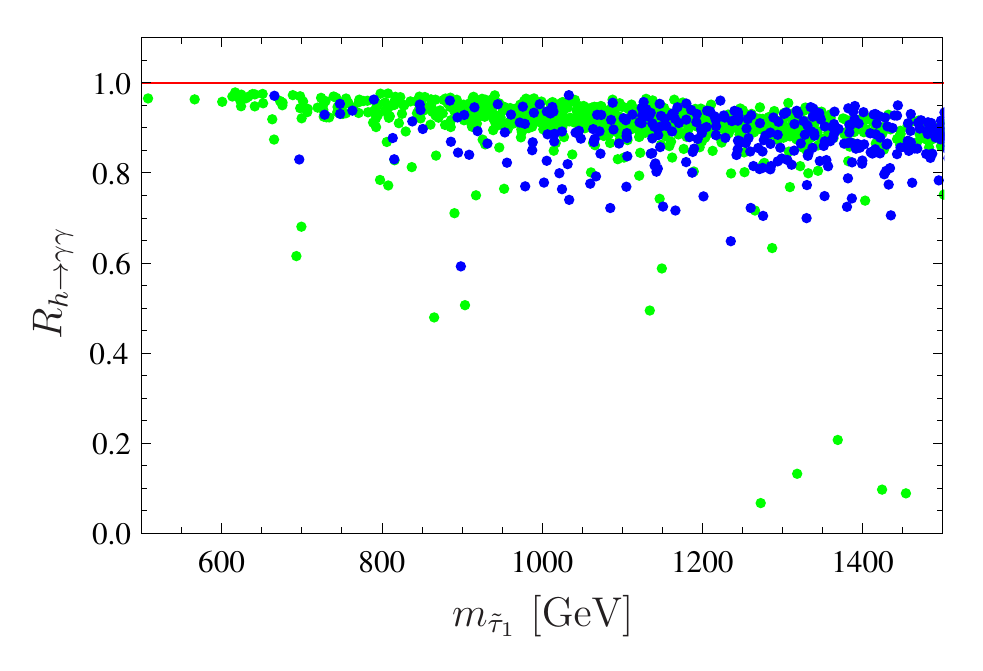}
 \caption{Decay rate of $h \to \gamma \gamma$ as a
 function of the stau mass using four \textbf{10}-plets. Only points with $123~\text{GeV} < m_h < 128~\text{GeV}$ are included.
 Color coding of the parameter points is as in \FIG \ref{fig:higgs_stop_scatter}.}
 \label{fig:stau_brgg_scatter}
\end{figure}

In supersymmetric models,
an enhancement can be achieved via a light stau. In models with extended
gauge structures, such a light stau and thus an enhancement of the 
$\gamma\gamma$ rate can be obtained even in scenarios
with large soft SUSY breaking parameters  \cite{Basso:2012tr}, as
there are large negative contributions due to the $D$ terms of the extra
$U(1)$ to the stau mass. However, in the model considered here, this does
not work for two reasons: the large stop mass required to obtain
the correct Higgs mass implies a lower limit on $\Lambda$, and, secondly
the $D$ term itself is smaller in our model compared to the one
of Ref.~\cite{Basso:2012tr} taking the same $Z'$ mass 
and ratio of additional \vevs
as demonstrated in Fig.~\ref{fig:Dterms}. 
In this restricted
model, the tadpole equations imply that a larger $v_R$ requires a smaller
$\tan \beta_R$, and thus the $D$ terms cannot be enhanced to the
required level.
In \FIG \ref{fig:stau_brgg_scatter}, we show $R_{h \to \gamma \gamma}$ as a 
function of the stau mass, demanding that the mass of the doubletlike Higgs 
to be in the range $123~\text{GeV} < m_h < 128~\text{GeV}$. 
This implies  a lower limit
on $m_{\tilde \tau_1} \gsim 500$~GeV, which is too large to get
a sizable contribution to $h\to\gamma\gamma$, and thus we find
 $R_{h \to \gamma \gamma} \lsim 1$ in this model. Hence, this model will be excluded 
 if $R_{h\to \gamma\gamma} > 1$
 is established by ATLAS and CMS at a significant level. However, the 
 most recent
 results of CMS point exactly in this direction that the diphoton rate is in good
 agreement with SM expectations \cite{CMS:diphoton}.

\subsection{Next-to-lightest SUSY particle}
As usual in GMSB models, 
the gravitino $\tilde G$ is the LSP, and 
its mass is given by 
\cite{Giudice:1998bp} 
\begin{equation}
 m_{3/2} = \frac{F}{\sqrt{3} m_{Pl}}\,,
\end{equation}
with the reduced Planck mass $m_{Pl}$. The gravitino mass is usually in the MeV 
range or above due to large messenger scales required to obtain
the correct symmetry breaking.
In our model are three possibilities for the next-to lightest supersymmetric 
particle (NLSP): the two usual candidates, which are
the lightest neutralino and the lightest slepton, which is usually a stau, and, 
in addition, our model contains the lightest sneutrino as the third candidate. 
In general, we can state that the lightest neutralino will be the NLSP for low 
messenger multiplicities, $n \lesssim 2$, and little hierarchy 
in the diagonal entries of $Y_S$.
For larger $n$, $\tilde \chi^0_1$ can only
be lighter than the lightest slepton if the left-right-splitting of 
the stau is small (\IE for low $\tan \beta$ values) 
or if $|\mu_R|$ is small. We  discuss the different character of a 
neutralino NLSP in the next Sec. \ref{subsec:neutralinoNLSP}.
Otherwise, \IE, for large $n$ and nonhierarchical $Y_S$, the stau  is the NLSP. A sneutrino can be the NLSP for all $n$ if there is a large
hierarchy in the $Y_S$ entries:  the scalar singlet field corresponding 
to the smallest $Y_S$ entry gets light.
In the following, we present the 
corresponding mass matrices and discuss briefly the main differences compared 
to the phenomenology of the usual minimal GMSB model using the parameter points
in \TAB \ref{tab:parameter_points}.
As the lifetime of the NLSP
is proportional to $F^2$ \cite{Giudice:1998bp}, we find that, in most of
the available parameter space, the NLSP is so long-lived that it will leave
a typical collider detector before decaying. However, its lifetime is,
in general, still below the bounds set by big bang nucleosynthesis.

\subsubsection{Neutralinos}
\label{subsec:neutralinoNLSP}
This model contains seven neutralinos, which are, beside the usual
MSSM gauginos and Higgsinos, the extra $U(1)$ gaugino $\lambda_\chi$ and the
two  $R$ Higgsinos $\tilde \chi_R$ and $\tilde{\bar \chi}_R$. In the basis 
$(\lambda_{Y},\lambda_{W^3},\tilde h^0_d, \tilde h^0_u,\lambda_\chi,\tilde{\bar \chi}_R,\tilde \chi_R)$,
the mass matrix reads
\begin{widetext}
\begin{align}
M_{\tilde \chi^0} = 
\begin{pmatrix}
                      M_{1} & 0  & -\frac{\gy v_d}{2} & \frac{\gy v_u}{2} & \frac{M_{Y\chi}}{2} &
                      0 & 0\\
                      0 & M_{2} & \frac{g_L v_d}{2} & -\frac{g_L v_u}{2} & 0 & 0 & 0\\
                      -\frac{\gy v_d}{2} & \frac{g_L v_d}{2} & 0 & -\mu & \frac{(\gchi-\gychi) v_d}{2} & 0 & 0\\
                      \frac{\gy v_u}{2} & -\frac{g_L v_u}{2} & -\mu & 0 &-\frac{(\gchi-\gychi) v_u}{2} & 0 & 0\\
                      \frac{M_{Y\chi}}{2} & 0 & \frac{(\gchi-\gychi) v_d}{2} & -\frac{(\gchi-\gychi) v_u}{2}& M_{\chi} & 
                      \frac{5 \gchi v_{\bar \chi_R}}{4}& -\frac{5 \gchi v_{\chi_R}}{4} \\
                      0 & 0 & 0 & 0 & \frac{5 \gchi v_{\bar \chi_R}}{4}
                      & 0 & -\mu_R \\
                      0 & 0 & 0 & 0 & 
                      -\frac{5 \gchi v_{\chi_R}}{4}& -\mu_R & 0 
                     \end{pmatrix}\,.                     
\end{align}
\end{widetext}

\begin{figure}[t]
 \includegraphics[width=\linewidth]{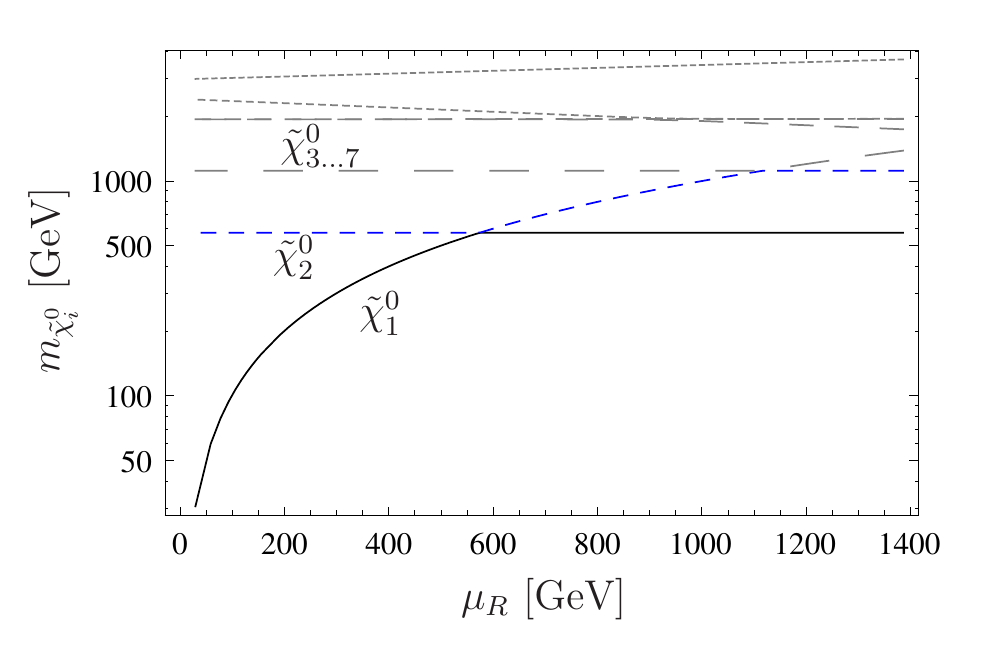}
 \caption{Neutralino masses 
 as a function of $\mu_R$ for the
 point BLRIV ($M_1\simeq 575$~GeV) and  $1.02 \le \tan\beta_R \le 1.033$ to 
 satisfy the tadpole equation (\ref{eq:tadpole_MuR}). 
}
 \label{fig:neutralinos_MuR}
\end{figure}

For a first understanding, it is useful to  neglect the mixing between the
MSSM states and the additional ones. In this case, one gets
$\mzp^2 \simeq \frac{25}{16} g^2_\chi v_R^2$, and, in the limit $\tan \beta_R \to 1$, one
finds for the eigenvalues of the three additional neutralino states
\begin{align}
\mu_R\,, \hspace{.2cm} \frac{1}{2}\left( M_\chi + \mu_R \pm \sqrt{\frac{1}{4} \mzp^2  + M_\chi^2 - 2 M_\chi \mu_R + \mu_R^2}\right) \, .
\end{align} 
In most of the parameter space, one finds $|\mu_R|,M_\chi \ll \mzp$, and thus
one has one state with  mass $|\mu_R|$ and two states with masses
close to $\mzp$, which can even form a quasi-Dirac state. For the MSSM-like
states, the lightest one is always binolike, and thus we find, depending
on the ratio $|\mu_R|/M_1$, that the lightest neutralino is either binolike
or a nearly maximal mixed $\tilde{\bar \chi}_R-\tilde \chi_R$ state.
This is exemplified in \FIG \ref{fig:neutralinos_MuR} for the point BLRIV
with a slight adjustment of $\tan\beta_R$ to satisfy the tadpole equation (\ref{eq:tadpole_MuR}).
Here, the NLSP nature changes from $\tilde \chi_R$-like to binolike at about 
$\mu_R \simeq 575$~GeV.

\begin{figure}
\centering
 \begin{minipage}{\linewidth}
 \includegraphics[width=\linewidth]{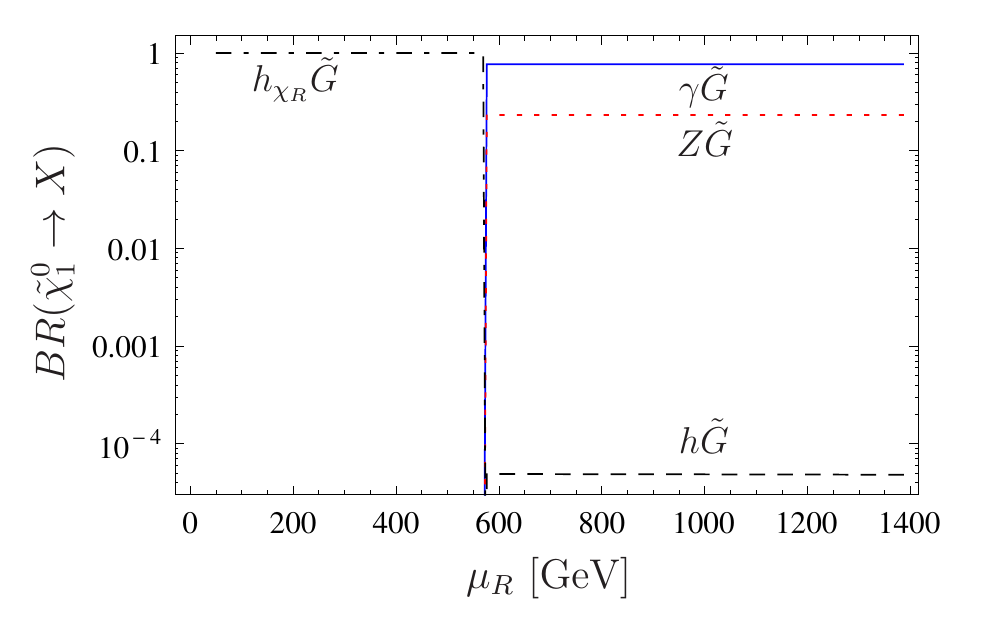}
 \end{minipage}
 \begin{minipage}{\linewidth}
 \includegraphics[width=\linewidth]{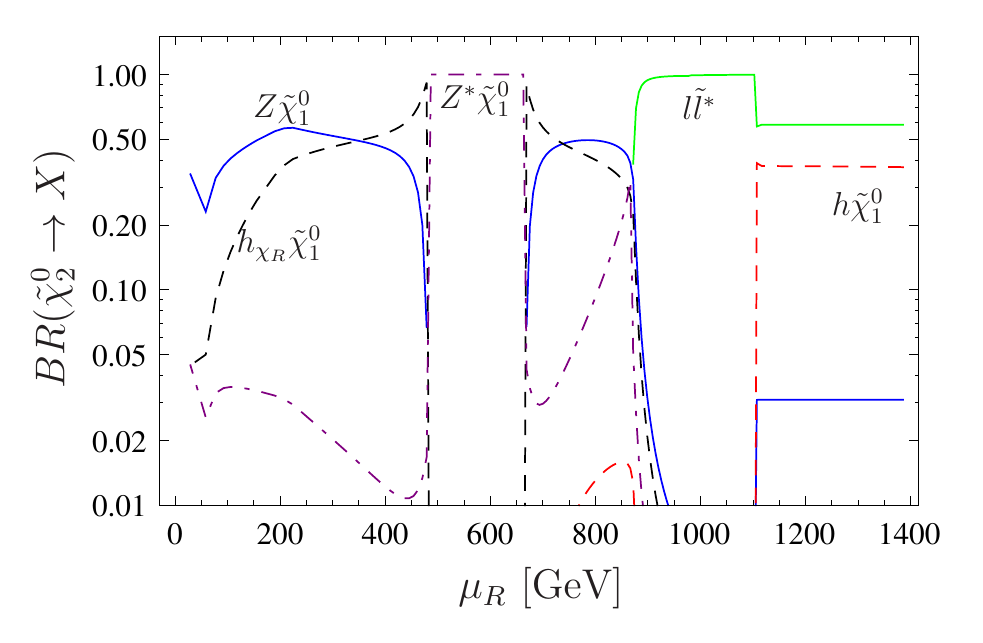}
 \end{minipage}
 \caption{Branching ratios of  $\tilde \chi^0_1$ (above)
 and of $\tilde \chi^0_2$ (below) as a function of $\mu_R$
 for the same parameter choice as in \FIG \ref{fig:neutralinos_MuR}.}
 \label{fig:chi_decays}
\end{figure}

The lightest neutralino will decay dominantly into a
$\chi_R$-like Higgs state and a gravitino $\tilde G$ if it is mainly
a $\tilde \chi_R$ Higgsino, whereas the MSSM-like bino state decays
dominantly into $\gamma\tilde G$ and $Z \tilde G$  as depicted in 
\FIG \ref{fig:chi_decays}. However, as mentioned above, the
neutralinos are rather long-lived, and thus, at the LHC, they will decay, in
general, outside the detectors. 
Hence, new techniques would be necessary to observe these states.
For $|\mu_R| < M_1$, we find that $h_{\chi_R}$ can be produced in the decays of
$\tilde \chi^0_2$ at a sizable rate. Therefore, SUSY cascade decays offer the
possibility to study this particle, which can hardly be produced directly or
in Higgs decays.

The large values of $\Lambda$ imply that the squarks and the gluino
are usually in the multi-TeV range, implying that one will need the high
luminosity option of the LHC to study these particles in detail. It turns out
that the two lightest states are $\tilde g$ and $\tilde t_1$.  Depending on
$m_{\tilde g} - m_{\tilde t_1}$, the gluino decays either dominantly
into third-generation quarks and neutralinos/charginos or into $t\tilde t_1$. 
In both cases, the final states will contain $b$ jets and $W$ bosons.
Depending on the nature of the two lightest neutralinos, also a Higgs boson
can be in the final state as discussed above.
Moreover, also, the additional sneutrinos can appear in the cascade decays,
but distinguishing them from the usual MSSM sneutrinos will be rather difficult.

\subsubsection{Charged sleptons}
\label{subsec:sleptonNLSP}

The mass matrix of the sleptons reads in the basis 
$\left(\tilde{e}_L, \tilde{e}_R \right)$
\begin{align}
 \notag &m^2_{\tilde l} = \\
 &\begin{pmatrix}
 m^2_L +\frac12 v^2 c^2_\beta Y_e^\dagger Y_e  + D_L {\bf 1}
  & \frac{v}{\sqrt 2} (T_e^\dagger c_\beta - \mu Y_e^\dagger s_\beta)\\
  \frac{v}{\sqrt 2} (T_e c_\beta - \mu^* Y_e s_\beta) & 
  m^2_E +\frac12 v^2 c^2_\beta Y_e Y_e^\dagger  + D_R {\bf 1}
 \end{pmatrix}\,,
\end{align}
which has the same structure as in the MSSM, but, for the concrete form
of the $D$ terms,
\begin{align}
\notag D_L =& \frac{1}{32} \Big(2\big( 
-3 \gchi^2 + \gchi \gychi + 2 (\gy^2 -g_L^2 +\gychi^2)
\big)v^2 c_{2\beta} \\
&- 5 \gchi(3 \gchi+2\gychi)v_R^2 c_{2 \beta_R}\Big) \,,\\
\notag D_R =&  \frac{1}{32} \Big(2\big( 
\gchi^2 + 3 \gchi \gychi - 4 (\gy^2 +\gychi^2)
\big)v^2 c_{2\beta}\\ &+ 5 \gchi( \gchi+4\gychi)v_R^2 c_{2 \beta_R}\Big)  \,.
\end{align}

\begin{center}
\begin{figure*}[t]
 \begin{minipage}{.49\linewidth}
 \includegraphics[width=\linewidth]{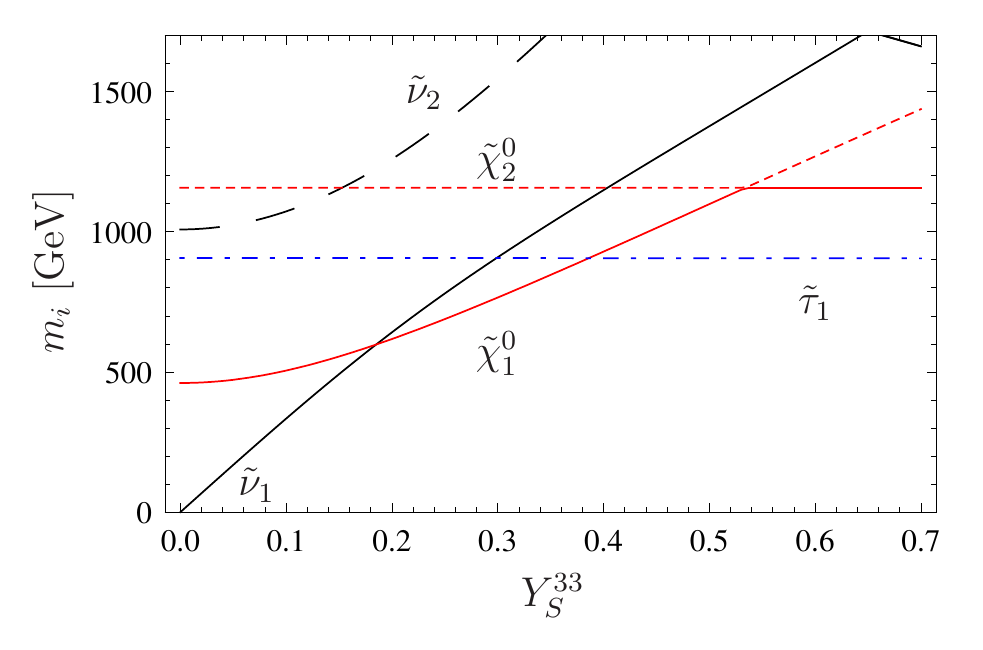}
 \end{minipage}
 \begin{minipage}{.49\linewidth}
 \includegraphics[width=\linewidth]{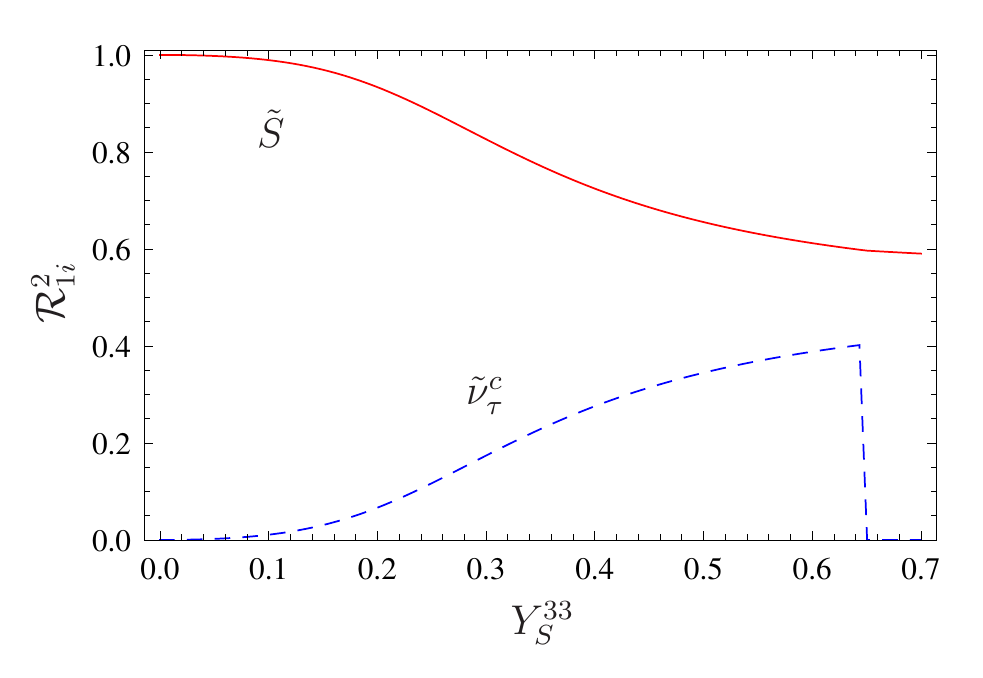}
 \end{minipage}
 \caption{Masses of the lightest SUSY particles (left) and 
 the sneutrino composition (right) 
 as a function of 
 $Y_S^{33}$ for the parameters specifying the
 points BLRI, BLRII, and BLRIII. In the right plot, the full red (dashed blue) line
 gives the $\tilde S$ ($\tilde \nu^c_{\tau}$) contribution to the nature
 of the lighest sneutrino.
}
 \label{fig:sneu_NLSP}
\end{figure*}
\end{center}

As explained above, stau NLSPs can be obtained for $n \ge 3$ and large values of 
$\tan \beta$ as  left-right mixing can compensate the additional $D$ terms.
As can be seen in \FIG~\ref{fig:sneu_NLSP}, the three diagonal
entries of $Y_S$ have to be of roughly the same size.

As indicated in \TAB~\ref{tab:parameter_points}, the gluino is usually
very heavy, and it turns out that it is the heaviest strongly interacting particle.
This implies that one will need a very high luminosity to discover this particle.
In general, it decays into all squarks, which in turn decay further into
the MSSM-like neutralinos and charginos. Note, that both light Higgs states,
the doubletlike one as well as the $h_{\chi_R}$-like one,
can be produced in these decays. Finally, the lightest neutralino will
decay into $\tau \tilde\tau_1$, and $\tilde \tau$ will, in general, decay outside
the detector. Thus, a typical event will consist of several jets and leptons
plus a charged track from a (at the detector level) stable particle. 
 The
 phenomenology of long-lived staus has already been studied comprehensively in the literature;
 see, \EG, Refs. \cite{Feng:1997zr,Drees:1990yw}, and bounds of $m_{\tilde \tau} \gtrsim 300~$GeV have been set by the LHC collaborations \cite{Aad:2012pra} in 
 MSSM scenarios.

\subsubsection{Sneutrinos}
\label{subsec:sneutrinoNLSP}

As it is well-known, in inverse seesaw scenarios, the parameter
$\mu_S$ has to be small to explain correctly neutrino data\footnote{For a discussion
and the corresponding  mass matrix, see, for example, Ref. \cite{Hirsch:2012kv} and references
therein.}. For completeness, we note that
the inverse seesaw mechanism yields in this model three very light Majorana states,
which can explain the observed neutrino data where the six heavier neutrinos 
are pairwise degenerate forming three quasi-Dirac states   \cite{Hirsch:2012kv}.
We will denote the former by $\nu$ and the latter by $\nu_h$.
The $F$ terms induced by $\mu_S$
as well as the corresponding soft SUSY breaking term $B_{\mu_S}\tilde S\tilde S$
induce a splitting of the complex sneutrino fields into their scalar and
pseudoscalar components. However, in practice, this mass splitting
is tiny, and thus we can safely neglect it in the following discussion. In the limit
$\mu_S, B_{\mu_S} \to 0$, 
the sneutrino mass matrix reads in the basis ($\tilde \nu,~\tilde \nu^c, \tilde S$)
\begin{widetext}
\begin{align}
 &M^2_{\tilde \nu} =\\ \notag &
  \begin{pmatrix}
 m_L^2 + \frac{v^2  s_{\beta}^2}{2} Y_\nu^\dagger Y_\nu +D_L' {\bf 1}
  & \frac{v}{\sqrt{2}} \big( T_\nu^\dagger s_{\beta} -\mu Y_\nu^\dagger c_{\beta} \big)  
                     &  \frac{1}{2}v v_R Y_\nu^\dagger Y_S s_{\beta} s_{\beta_R}\\
  \frac{v}{\sqrt{2}} \big( T_\nu s_{\beta} -\mu^* Y_\nu c_{\beta} \big) &
m^2_{\nu^c} + \frac{v_{R}^2  s^2_{\beta_R}}{2} Y_S Y_S^\dagger 
+ \frac{v^2 s_{\beta}^2 }{2} Y_\nu Y_\nu^\dagger +    D_R'   {\bf 1}  &
 \frac{v_R}{\sqrt{2}}\big(T_S s_{\beta_R} - \mu_R^* Y_S c_{\beta_R} \big) \\
                     \frac{1}{2}v v_R Y_S^\dagger Y_\nu s_{\beta} s_{\beta_R} 
                     & \frac{v_R}{\sqrt{2}}\big(T_S^\dagger s_{\beta_R} - \mu_R Y_S^\dagger c_{\beta_R} \big)
                     & ~m^2_S + \frac{v_R^2 s_{\beta_R}^2}{2} Y_S^\dagger Y_S 
                    \end{pmatrix}\,,
                    \label{eq:sneutrino_massmatrix}
\end{align}
with
\begin{align}
D_L' =& \frac{1}{32} 
\Big(2 \big( -3\gchi^2 + \gchi \gychi + 2(g_L^2 +\gy^2 +\gychi^2) \big) v^2 c_{2 \beta}  - 
 5 \gchi (3 \gchi + 2 \gychi) v_R^2 c_{2 \beta_R} \Big)  \,, \\ 
 D_R' =&  
 \frac{5 \gchi}{32} \Big( 2 (\gchi - \gychi) v^2 c_{2 \beta} 
 + 5 \gchi v_R^2 c_{2 \beta_R} 
 \Big) \,.
\end{align}
\end{widetext}

Obviously, the masses of the sneutrinos depend strongly on $Y_S$. In particular, $\tilde S$-$\tilde S$
entries are dominated by $\frac{v_R^2 s_{\beta_R}^2}{2} Y_S^\dagger Y_S$ because  $m_S^2$  is rather
small, as discussed in \sect{subsec:gmsb_boundaries}.
Therefore, even $m_S^2 < 0$ does not  automatically imply spontaneous $R$-parity breaking. 
The $\tilde \nu^c$-$\tilde S$ mixing entry can be of the same size as the corresponding diagonal entries
for sufficiently large $|\mu_R|$. The entries which mix these states with $\tilde \nu_L$ are much smaller
and can be neglected for the moment. As $\tan\beta_R$ is close to 1, we can take the limits $\tan\beta_R\to 1$,
$D_R' \to 0$
and find for these approximations the upper bound,
\begin{equation}
|\mu_R| \lsim \sqrt{m^2_{\nu^c} +v_R^2 Y_S^2/4}\,,
\end{equation}
to avoid tachyonic states. Here, we have also set $T_S=0$, as this is numerically always small.
For completeness, we note that $|Y_S Y^\dagger_S|$ is bounded from above by the requirement 
that all couplings stay perturbative up to the GUT scale and from below by the requirement
of correct symmetry breaking as discussed in \sect{subsec:tadpoles}.

Combining all requirements, we find that a light sneutrino, which could be the  NLSP, if one of 
the diagonal 
$Y_S$ entries is rather small, $\lsim 0.2$, and the other two are large, $\sim 0.7$. In an abuse
of  language, we call this state a sneutrino, even though
the corresponding state is dominantly a $\tilde S$. However, it still can have a sizable $\tilde \nu^c$
admixture as exemplified  in  \FIG~\ref{fig:sneu_NLSP}. Note that taking $Y^{33}_S$ small is an
arbitrary choice, and we could have equally well taken one of the two other generations. The smallness
of this coupling also implies that one of the heavy quasi-Dirac neutrinos is significantly lighter than the other
two, but we find that this state is always heavier than the lightest sneutrino. Therefore, a sneutrino NLSP
decays always invisibly into $\nu \tilde G$. As can be seen in \FIG~\ref{fig:sneu_NLSP}, the next heavier
state is $\tilde \chi^0_1$, which turns out to be mainly a $\tilde \chi_R$ Higgsino. If kinematically
allowed, it will decay to $\tilde \nu_1 \nu_h$, yielding
\begin{equation}
\tilde \chi^0_1 \to \tilde \nu_1 \nu_h \to \nu \tilde G W^{(*)} l\,,
\end{equation}
giving a final state with an (off-shell) $W$ boson, the lepton of the corresponding generation and missing
energy. For completeness, we note that we find $BR(\tilde \chi^0_1) \to \tilde \nu_1 \nu_h \simeq 1$ if
$|Y_S^{33}|\lsim 0.07$ for the parameters used in \FIG \ref{fig:sneu_NLSP} and  
$BR(\tilde \chi^0_1) \to \tilde \nu_1 \nu \simeq 1$ for larger values of $|Y_S^{33}|$. The latter
leads to a completely invisible final state, and thus, in this part of the parameter space,
this scenario cannot be distinguished from the $\tilde \chi^0_1$ NLSP case in this model.


\subsection{$Z'$ phenomenology}
\label{subsec:zp_exclusion_bounds}
\begin{figure}
 \includegraphics[width=\linewidth]{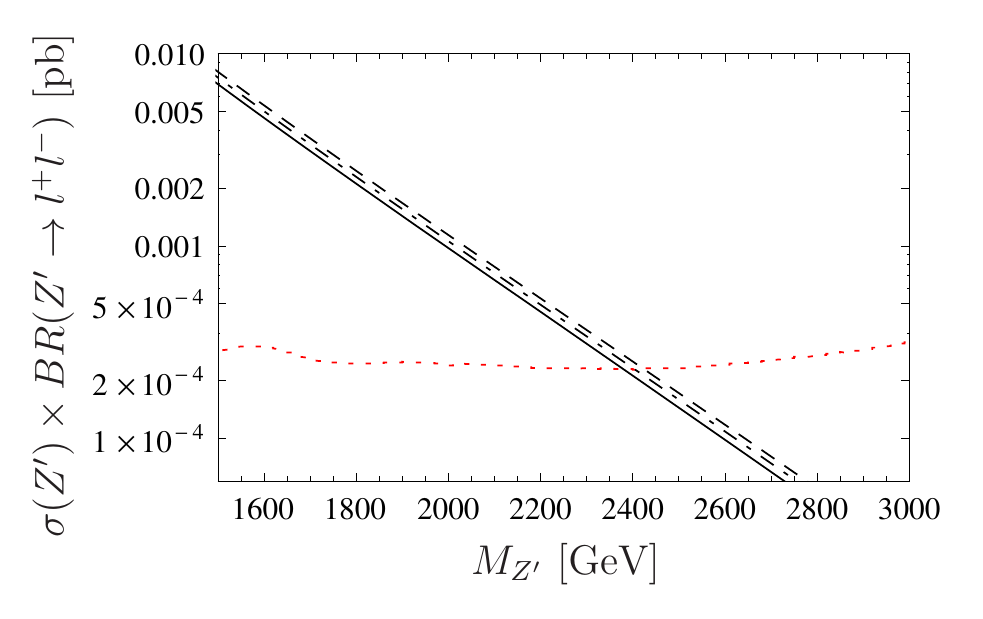}
 \caption{$\sigma(pp \to Z' \to l^+ l^-)$ as a function of $\mzp$ for the scenarios 
 BLRI (solid line), BLRII (dotted-dashed line), and BLRIII (dashed line).
 The red dotted line shows the exclusion limits at 95~\% C.L. obtained by ATLAS \cite{ATLAS-CONF-2013-017}. 
 }
 \label{fig:zp_bounds}
\end{figure}

The LHC collaborations ATLAS and CMS have recently updated the bounds on $\mzp$  from the search
for dilepton resonances \cite{ATLAS-CONF-2013-017, CMS-PAS-EXO-12-061} at  $\sqrt s = 8~$TeV and an integrated 
luminosity of about 20~fb$^{-1}$ each. In order 
to apply these bounds to our model we calculate the production cross section of the $Z'$ and the subsequent 
decay into a pair of leptons as a function of the $Z'$ mass\footnote{We used {\tt CalcHEP 3.4.2} \cite{Belyaev:2012qa} for the cross section
calculation. The model was implemented using the {\tt SUSY Toolbox} \cite{Staub:2011dp}.}. 
From \FIG~\ref{fig:zp_bounds}, one can extract the limits that depend on the underlying
parameters. In case of BLRIII (dashed line),
only standard model decay channels for the  $Z'$ are open, leading to a bound of about 2.43~TeV,
whereas it can be reduced to about 2.37~TeV if, in addtion, decays into heavy neutrinos and
sneutrinos are allowed as is the case of BLRI (full line). 
 This translates into a lower limit 
on  $v_R$ of about $v_R \gtrsim 6.6~$~TeV.

As already mentioned, the colored SUSY particles are rather heavy in this model 
in most of the 
parameter space once the constraint on the Higgs mass is imposed
implying that the discovery of supersymmetry requires
either a huge statistics and/or a larger c.m. 
energy. However, it has been shown that the decays of the $Z'$ open the
possibility to produce  SUSY particles 
\cite{Chang:2011be,Krauss:2012ku,Hirsch:2012kv,Corcella:2012dw,Gherghetta:1996yr,Kang:2004bz,Baumgart:2006pa}. 
As has been discussed in Ref. \cite{Hirsch:2012kv}
in a constrained-MSSM-like variant of this model, the potentially interesting
final states from $Z'$ decays are: $\nu_h \nu_h$, $\tilde l\tilde l$, 
$\tilde \nu\tilde \nu$, $\tilde \chi^+_i \tilde \chi^-_i$, and
$\tilde \chi^0_i \tilde \chi^0_j$.   However, it turns out that, in
the GMSB variant, the required conditions for the different channels 
are harder to realize, as this model is more constrained. In particular, we hardly find
charged sleptons except for the case that $\mzp$ is above 4 TeV because it couples
significantly stronger to $L$ sleptons than to $R$ sleptons.
In \TAB \ref{tab:Zp_decays}, 
we list the $Z'$ decay modes of
the parameter points given in \TAB \ref{tab:parameter_points}.

\begin{table}[t]
\centering
\begin{tabular}{| l | c | c | c | c | c | c |} \hline \hline
  & BLRI&BLRII&BLRIII&BLRIV&BLRV&BLRVI\\ \hline\hline
  $\mzp$ [TeV] & \multicolumn{3}{|c|}{2.5}  & 2.7 & 2.4 & 4.3 \\ \hline
  $BR(d\bar d)$ & 0.45 & 0.49 & 0.52 & 0.52 & 0.52 & 0.48\\
  $BR(u\bar u)$ & 0.08 & 0.09 & 0.10 & 0.10 & 0.10 & 0.09\\
  $BR(l\bar l)$ & 0.17 & 0.18 & 0.20 & 0.20 & 0.20 & 0.18\\
  $BR(\nu \nu)$ & 0.15 & 0.16 & 0.17 & 0.17 & 0.17 & 0.16\\
  $BR(W^+ W^-)$ & 0.01 & 0.01 & 0.01 & 0.01 & 0.01 & 0.01\\   
  $BR(\nu_h \nu_h)$ & 0.12 & 0.06 & $-$ & $-$ & $-$ & $-$\\
  $BR(h_1 Z)$ & $-$ & $-$ & 0.01 & $-$ & $-$ & $-$\\
  $BR(h_2 Z)$& $-$ & $-$ & $-$ & 0.01 & $-$ & $-$\\
  $BR(\tilde l \tilde l^*)$ & $-$ & $-$ & $-$ & $-$ & $-$ & 0.02\\
  $BR(\tilde \nu \tilde \nu)$ & 0.01 & $-$ & $-$ & $-$ & $-$ & 0.01\\ 
  $BR(\tilde \chi^0_i \tilde \chi^0_j)$ & $-$ & $-$ & $-$ & $-$ & $-$ & 0.02 \\
  $BR(\tilde \chi^+_2 \tilde \chi^-_2)$ & $-$ & $-$ & $-$ & $-$ & $-$ & 0.02 \\
  \hline
  \hline
\end{tabular}
\caption{Branching ratios of the $Z'$ boson for the parameter points of \TAB \ref{tab:parameter_points}. 
Only branching ratios  larger than $10^{-2}$ are shown.}
\label{tab:Zp_decays}
\end{table}

The most important nonstandard decays of $Z'$ are those into heavy neutrinos. Their masses are
proportional to $\sqrt{Y_\nu^2+Y_S^2} v_R$, implying that the corresponding Yukawas should not be too large
because otherwise these decays are kinematically forbidden. This can clearly be seen 
by combining Tables \ref{tab:parameter_points} and \ref{tab:Zp_decays}: the smaller the $Y_S$, the larger
the corresponding branching is (for fixed $Y_\nu$). The heavy neutrinos decay into $W l$, $Z \nu$
and $h \nu$ with a branching ratio of $\sim 0.6$, $\sim 0.2$, and $\sim 0.2$, respectively 
\cite{Hirsch:2012kv}. Here, $h$ denotes the doubletlike Higgs boson.

Naively, one would expect that also sneutrinos
should show up in such scenarios because, as discussed in \sect{subsec:sneutrinoNLSP},
the smaller the $Y_S$, the smaller the mass of the lightest sneutrino. However, at the same time, 
the $\tilde S$ component increases, as can be seen in \FIG~\ref{fig:sneu_NLSP}, which reduces the coupling
to the $Z'$. For an intermediate range of $Y_S$, the second lightest sneutrino can be produced
in $Z'$ decays. It decays dominantly according to
\begin{align}
 \tilde \nu_2 \to \nu_h \tilde \chi^0_1 \to \nu_h \nu_h \tilde \nu_1 \to l l  W W + E\!\!\!/_T\,,
\end{align}
yielding
\begin{equation}
Z' \to \tilde \nu_2 \tilde \nu_2^* \to 4 l 4 W E\!\!\!/_T
\end{equation}
as a final state. The other final states are $2 l 2 W 2 Z E\!\!\!/_T$, $2 l 2 W 2 h E\!\!\!/_T$,
$4 Z E\!\!\!/_T$,$2 Z 2 h E\!\!\!/_T$, and $4 h E\!\!\!/_T$, where $h$ denotes again the doubletlike
Higgs boson. We note for completeness that, for some part of the parameter space, also
the decay $\tilde \nu_2 \to \tilde \nu_1 h_{\chi_R}$ is possible offering, in principle, a 
possiblity to observe $h_{\chi_R}$. 
As $\tilde \nu_2$ is relatively heavy, there is a kinematical suppression and 
 we find only branching ratios of at most $O(0.01)$ for sneutrinos in
the final state.

The $Z'$ couples to all Higgsino states and the correspond coupling is proportional to
\begin{equation}
\gchi \left(2 (Z_\chi^{i,3}Z_\chi^{j,3}-Z_\chi^{i,4}Z_\chi^{j,4}) 
+ 5 (Z_\chi^{i,6}Z_\chi^{j,6}-Z_\chi^{i,7}Z_\chi^{j,7}) \right) \,.
\end{equation}
Here, $Z_\chi$ is the unitary matrix which
diagonalizes the neutralino mass matrix.
As discussed above, the $\tilde \chi_R$-like neutralinos can be
rather light. However, its admixture is such that $Z_\chi^{1,6} \simeq Z_\chi^{1,7} \simeq \pm 1/\sqrt{2}$,
and thus this final state has only a tiny branching ratio. For large $\mzp$, the decays into to the heavy MSSM
charginos/neutralinos containing a sizable Higgsino component are kinematically allowed with 
a branching ratio of a few percent as can be seen in \TAB~\ref{tab:Zp_decays}.
In this scenario, the production cross section for the $Z'$ is about 1 fb, and thus again large
statistics are needed to observe and study the corresponding final states.

\subsection{Lepton flavor violation}
\label{sec:LFV}
So far, we have assumed that neutrino mixing is explained by the flavor struture of $\mu_S$, which
hardly plays a role for the phenomenology discussed so far. In this case, also the rates for
flavor violating lepton decays are very small and cannot be observed in the near future.
However, as can be seen from the seesaw approximation of the neutrino mass matrix \cite{Mohapatra:1986bd},
\begin{align}
 m_\nu^{IS} \simeq \frac{v_u^2}{v^2_R} Y_\nu^T Y_S^{-1} \mu_S (Y_S^T)^{-1} Y_\nu\,,
\end{align}
neutrino mixing can also be explained by the flavor structure of $Y_\nu$ and $Y_S$. As one can always
find a basis where $Y_S$ is diagonal on the expense of having nondiagonal $Y_\nu$ and $\mu_S$, we will
now consider the other extreme case in which the complete flavor structure resides in $Y_\nu$.
Nondiagonal $Y_\nu$ induces also nondiagonal entries in the soft-breaking terms of sleptons. However,
as the scale for the GMSB boundary is lower than the GUT scale, the corresponding effects are significantly
smaller compared to typical SUGRA scenarios. Therefore, the main effect is due to the vertices for which
the off-diagonal entries of $Y_\nu$ enter.
To study the corresponding effects in our model, we parametrize $Y_\nu$  
according to  \cite{DeRomeri:2012qd}:
\begin{align}
\notag  Y_\nu =& f \begin{pmatrix}
          0 & 0 & 0 \\
          a & a (1-\frac{\sin \theta_{13}}{\sqrt{2}}) & - a (1+\frac{\sin \theta_{13}}{\sqrt{2}}) \\
          \sqrt{2} \sin \theta_{13} & 1 & 1
         \end{pmatrix} \label{eq:yv_parametrization}, \\
    &a = \left(\frac{\Delta m_{\odot}^2}{\Delta m_\text{Atm}^2}\right)^{\frac{1}{4}} \approx 0.4\, ,
\end{align}
using the latest data from the global fit of the PMNS matrix \cite{GonzalezGarcia:2012sz}.
This fixes $Y_\nu$ up to a global free prefactor $f$, which  determines the rate for flavor violating decays
 like $\mu \to e \gamma$ or $\mu \to 3e$. Their branching ratios 
are constrained by experiment to be smaller than $5.7\cdot 10^{-13}$ and $10^{-12}$, respectively \cite{Adam:2013wea, Beringer:1900zz}. 
In addition, the $\mu$-$e$ conversion rate (CR) in gold, which has to be smaller than $7 \cdot 10^{-13}$,  turns
out to be important  \cite{Bertl:2006up}.
\begin{center}
\begin{figure*}[t]
\centering
  \begin{minipage}{.49\linewidth}
   \includegraphics[width=\linewidth]{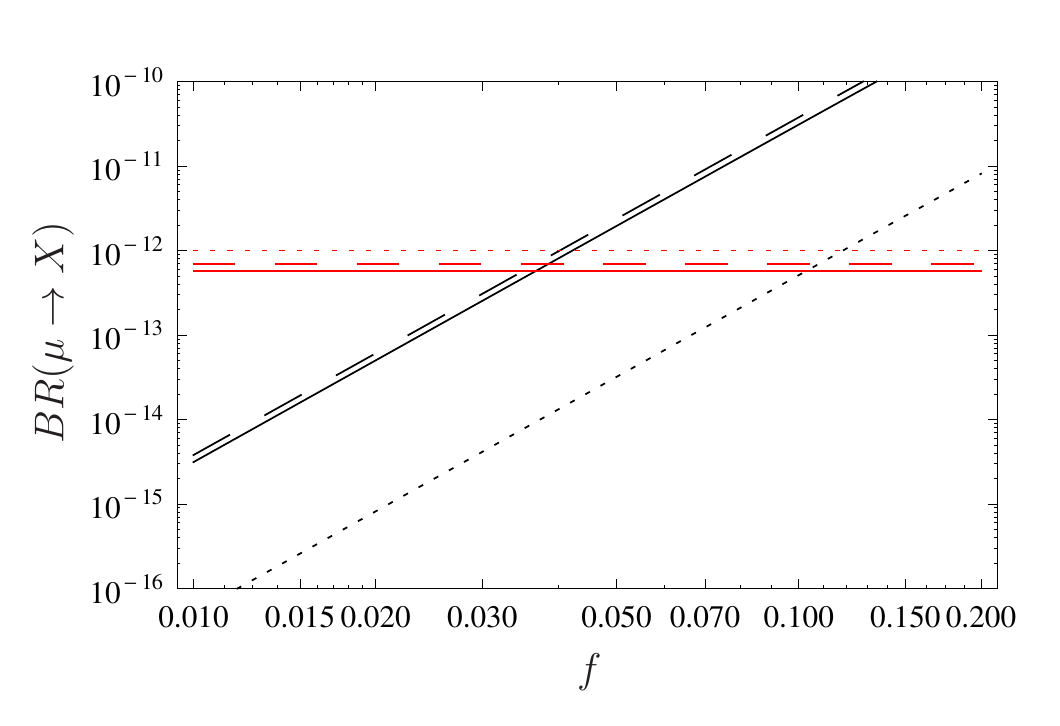}
   \end{minipage}
   \begin{minipage}{.49\linewidth}
   \includegraphics[width=\linewidth]{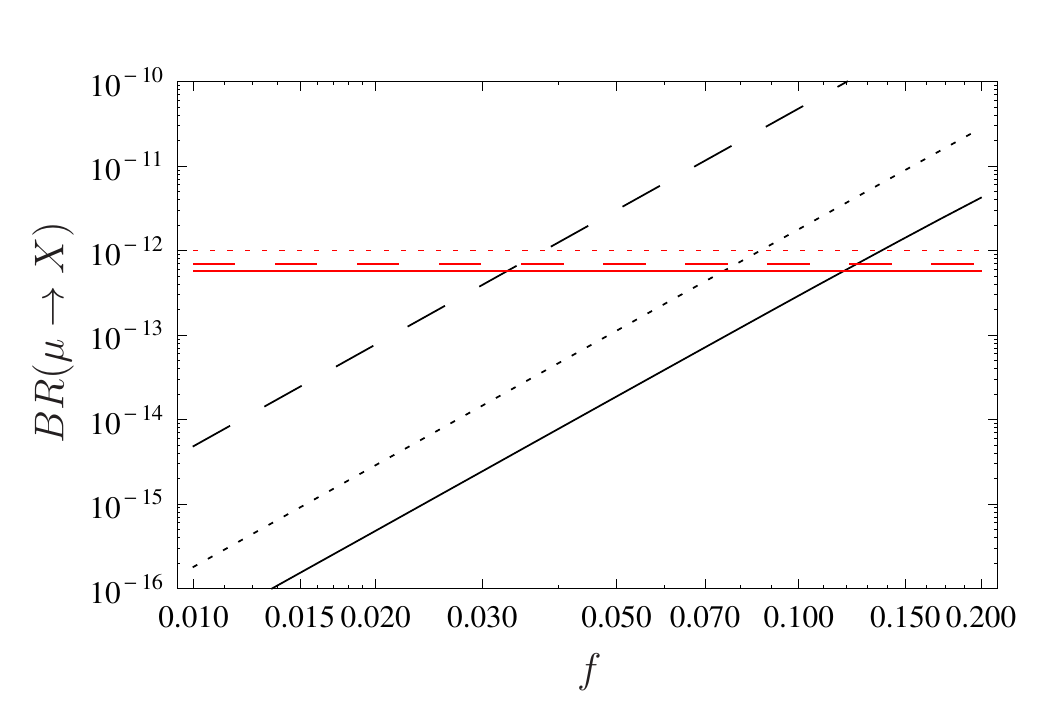}
   \end{minipage}
 \caption{Flavor violating observables as a function of $f$ as defined in \EQ~(\ref{eq:yv_parametrization}):
 $BR(\mu \to e \gamma$) (solid line), $BR(\mu \to 3 e)$ (dotted line) and  
 $CR(\mu \to e)$ in $ Au$ (dashed line) for the points 
 BLRI (left) and BLRIII (right) defined in \TAB~\ref{tab:parameter_points} and
 using $Y_\nu$ to explain the neutrino data.
 The upper bounds ($BR(\mu\to e\gamma) < 5.7 \cdot 10^{-13}$  \cite{Adam:2013wea},
 $BR(\mu\to 3e) < 1.0\cdot 10^{-12}$ \cite{Beringer:1900zz}, $CR(\mu-e,\mbox{Au}) < 7.0\cdot 10^{-13}$ 
 \cite{Bertl:2006up}) are shown as a red horizontal line, respectively. 
 } 
 \label{fig:LFV_standard_plots}
\end{figure*}
\end{center}

\begin{figure}[t]
 \centering
 \includegraphics[width=\linewidth]{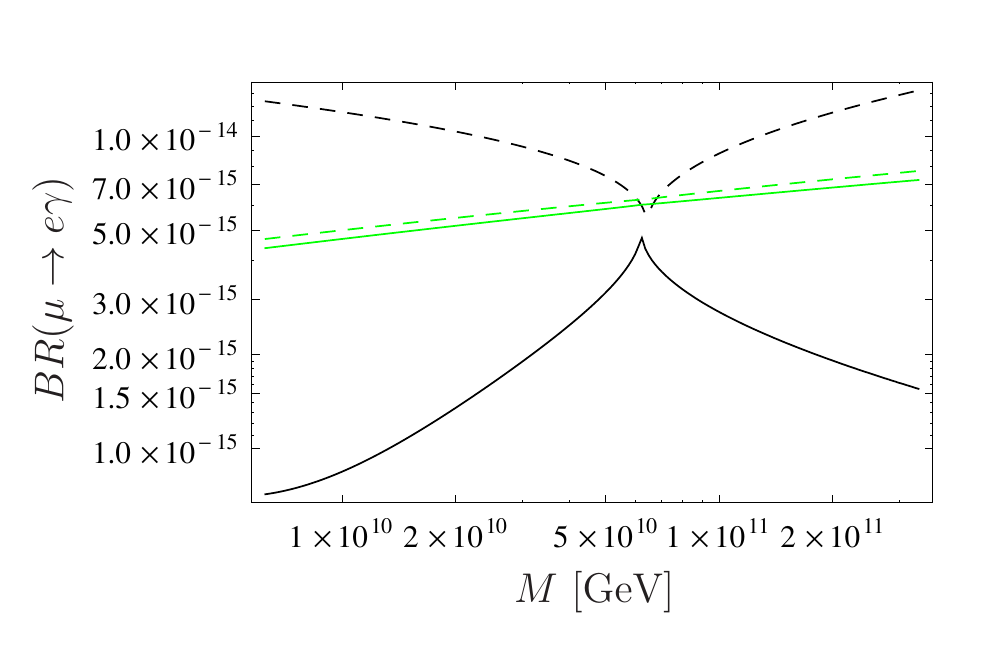}
 \caption{
 Branching ratios of the LFV decay $BR(\mu \to e \gamma)$ as a function of the messenger scale $M$ for
 sign $(\mu_R) = +$ (black solid line) and sign $(\mu_R) = -$ (black dashed line) fixing the other parameters
 as in BLRIV and $f=0.03$.
 The respective straight green (light) lines show the branching ratio excluding the contribution
 of  charged Higgsinos and sneutrinos.}
 \label{fig:LFV_scale}
\end{figure}

In Figure \ref{fig:LFV_standard_plots} we show these rates as a function of $f$ for the two points
BLRI and BLRIII. We observe that, in these scenarios $CR(\mu-e,\mbox{Au})$ is the most constraining
observable followed by $BR(\mu\to 3e)$ and/or $BR(\mu\to e\gamma)$. As explained in detail in 
Refs.~\cite{Hirsch:2012ax} and \cite{Abada:2012cq,Ilakovac:2012sh}, this behavior can be understood
as a nondecoupling effect of the $Z$-boson contribution to $CR(\mu-e,\mbox{Au})$ and 
$BR(\mu\to 3e)$, which are enhanced by a factor  $(m_{SUSY}/M_Z)^4$ with respect to the off-shell
photon contributions, which is sizable due to the required heavy SUSY spectrum.
 In all cases, the sneutrino-chargino loops give the dominant contributions.
We find that $CR(\mu-e,\mbox{Au})$ gives the strongest
constraint on the size of $f$  for all points of \TAB~\ref{tab:parameter_points}.
For completeness, we note that, once the bounds on this observable are fulfilled, we find
that the corresponding LFV decays of the $\tau$ are so suppressed that they are below
the reach of experiments in the near future. This implies, on the other hand, that, if, for
example, one of the LHC experiments observes, for example, $\tau\to 3 \mu$, then this class of
models is ruled out.   

A point worth mentioning here is a rather strong dependence of $BR(\mu \to e \gamma)$ on
the sign of $\mu_R$,  which can change the rate by 1 order of magnitude. The reason is
that the off-diagonal elements for the $\tilde \nu^c$-$\tilde S$ mixing in \EQ (\ref{eq:sneutrino_massmatrix})
are dominated by
the $\mu_R v_R$ contribution as discussed in \sect{subsec:sneutrinoNLSP}.
 We exemplify this behavior in \FIG~\ref{fig:LFV_scale}, where we show $BR(\mu \to e \gamma)$
as a function of $M$ for both signs of $\mu_R$, $f=0.03$ and fix the other parameters as for BLRIV.
The black lines give all contributions, whereas, for the light green lines, we have taken out the
contributions containing the Higgsino-like chargino. As a consequence, $BR(\mu \to e \gamma)$ could be
in the reach of an upgraded version of the MEG experiments if $\mu_R$ is negative.
For completeness, we note that this is a specific feature of the GMSB model as, for example, in SUGRA
inspired models, large trilinear couplings $T_S$ could be present dominating this mixing.
 
Finally, we stress again that the finding of this section depends on the assumption that the complete
flavor structure needed to explain neutrino data is present in $Y_\nu$. If this structure is at least partially shifted to
$\mu_S$, one can reduce the predictions for the lepton flavor violating observables. It turns out that
the size of this reduction depends on the SUSY parameters, and thus we do not discuss it here in detail.

\section{Conclusion}
\label{sec:conclusions}
We studied in this paper the GMSB variant of a SUSY model with an
extended gauge sector in which the couplings unify at the GUT scale.
Compared to GMSB with MSSM particle content only, one can obtain
a tree-level mass for the light doubletlike Higgs boson above 
$M_Z$, which eases the need for large radiative corrections to obtain 
a Higgs mass at 125~GeV. For this reason, we find in this model 
a lower bound on the mass of the lighter stop of about two TeV, which
is about a factor of 2 smaller than in usual GMSB models. Nevertheless,
the SUSY particles of the strongly interacting sector are rather heavy, 
and thus the existing bounds on squarks and gluinos are satisfied 
automatically. However, this implies that one needs high luminosity
at the LHC with $\sqrt{s}=14$~TeV to study this sector and the
resulting cascade decays in detail. The rather heavy SUSY spectrum also
implies that the rate for the doubletlike Higgs boson decaying
into two photons is always below or at most the SM expectation. 
Therefore, this model can be ruled out if this rate turns out to 
be larger than the SM expectation at a significant level. 

This model contains an additional candidate for the NLSP: besides
the lightest neutralino or one of the sleptons, also a sneutrino
can be the NLSP. For this to happen, the additional Yukawa
coupling $Y_S$ needs to have a hierarchical structure. Moreover,
the stau NLSP is somewhat more difficult to achieve than in usual GMSB
models. We have worked out main features of the corresponding
scenarios paying also particular attention to the possiblity that
the new Higgs boson, which can be rather light, can show up
in the SUSY cascade decays. We have argued that  $Z'$ decays
can serve as a SUSY discovery even in this
rather restricted model.

Last but not least, we have discussed which lepton flavor violating 
observables can be observed in this class of GMSB
models. It turns out that $\mu\to 3 e$ and
$\mu-e$-conversion are usually more constraining than 
$\mu\to e\gamma$. The necessary requirement for a possible
observation is that there are sizable
off-diagonal entries in the neutrino Yukawa coupling. It turns out 
that the rates for the corresponding $\tau$ decays is well below the
sensitivity once the contraints from the muon sector are taken into account.
Note, however, that the rates for all observables get tiny if neutrino
mixings are explained via the flavor structure of $\mu_S$ instead of the
flavor structure of $Y_\nu$.

\section*{Acknowledgments}

We thank Martin Hirsch for useful discussions.
This work has been supported  by  DFG Project \\No.\ PO-1337/3-1.

\begin{appendix}
\section{Mass matrices in the $U(1)_R \times \UBL$ basis}
Here, we give for completeness 
the mass matrices that were shown in the text for the original basis of $SU(3)_c \times SU(2)_L \times U(1)_R\times U(1)_{B-L}$.
\begin{widetext}
\subsection{Higgs mass matrix}
In the basis $(\sigma_d, \sigma_u, \bar \sigma_R, \sigma_R)$, the Higgs mass matrix is given by
\begin{align}
\notag &m^2_{h^0} = \\
&\begin{pmatrix}
\frac14 \tilde g_{LL}^2 v^2 c_\beta^2 +  m^2_A s_\beta^2 
& - \frac{s_{2 \beta}}{8}(4 m^2_A + \tilde g_{LL}^2 v^2) 
& \frac14 \tilde g^2 v v_R c_{\beta} c_{\beta_R}
& -\frac14 \tilde g^2 v v_R c_{\beta} s_{\beta_R}
\\
- \frac{s_{2 \beta}}{8}(4 m^2_A + \tilde g_{LL}^2 v^2) 
& \frac14  \tilde g_{LL}^2 v^2 s_\beta^2 +  m^2_A c_\beta^2 
& -\frac14 \tilde g^2 v v_R s_{\beta} c_{\beta_R}
& \frac14 \tilde g^2 v v_R s_{\beta} s_{\beta_R}
\\
\frac14 \tilde g^2 v v_R c_{\beta} c_{\beta_R}
& -\frac14 \tilde g^2 v v_R s_{\beta} c_{\beta_R}
& \frac14 \tilde g_{RR}^2 v_R^2 c_{\beta_R}^2 +  m^2_{A_R}s_{\beta_R}^2
& -\frac{s_{2 \beta_R}}{8} (4 m^2_{A_R}+\tilde g^2_{RR}v_R^2)  
\\
-\frac14 \tilde g^2 v v_R c_{\beta} s_{\beta_R}
& \frac14 \tilde g^2 v v_R s_{\beta} s_{\beta_R}
& -\frac{s_{2 \beta_R}}{8} (4 m^2_{A_R}+\tilde g^2_{RR}v_R^2) 
& \frac14 \tilde g_{RR}^2 v_R^2 s_{\beta_R}^2 +  m^2_{A_R}c_{\beta_R}^2
\end{pmatrix}\,,
\end{align}
where  ${\tilde g^2} = g_R(g_R-g_{BLR}) + g_{RBL}(g_{RBL}-g_{BL}),~
{\tilde g_{LL}^2} = g_L^2+g_R^2+g_{RBL}^2$, and ${\tilde g_{RR}^2} = (g_{BLR}-g_R)^2 +(g_{BL}-g_{RBL})^2$.

\subsection{Neutralino mass matrix}
The neutralino mass matrix in the basis 
$(\lambda_{B-L},\lambda_{W^3},\tilde h^0_d, \tilde h^0_u,\lambda_R,\tilde{\bar \chi}_R,\tilde \chi_R)$ reads
\begin{align}
 \notag &M_{\tilde \chi^0} = \\  
                     &\begin{pmatrix}
                      M_{B-L} & 0  & -\frac{g_{RBL} v_d}{2} & \frac{g_{RBL} v_u}{2} & M_{BLR} &
                      \frac{(g_{BL}-g_{RBL}) v_{\bar \chi_R}}{2}& -\frac{(g_{BL}-g_{RBL}) v_{\chi_R}}{2}\\
                      0 & M_{2} & \frac{g_L v_d}{2} & -\frac{g_L v_u}{2} & 0 & 0 & 0\\
                      -\frac{g_{RBL} v_d}{2} & \frac{g_L v_d}{2} & 0 & -\mu & -\frac{g_R v_d}{2} & 0 & 0\\
                      \frac{g_{RBL} v_u}{2} & -\frac{g_L v_u}{2} & -\mu & 0 &\frac{g_R v_u}{2} & 0 & 0\\
                      M_{BLR} & 0 & -\frac{g_R v_d}{2} & \frac{g_R v_u}{2} & M_{R} & 
                      -\frac{(g_R-g_{BLR}) v_{\bar \chi_R}}{2}& \frac{(g_R - g_{BLR}) v_{\chi_R}}{2} \\
                      \frac{(g_{BL}-g_{RBL}) v_{\bar \chi_R}}{2} & 0 & 0 & 0 & -\frac{(g_R-g_{BLR}) v_{\bar \chi_R}}{2}
                      & 0 & -\mu_R \\
                      -\frac{(g_{BL}-g_{RBL}) v_{\chi_R}}{2} & 0 & 0 & 0 & 
                      \frac{(g_R - g_{BLR}) v_{\chi_R}}{2} & -\mu_R & 0 
                     \end{pmatrix}\,.
\end{align}

\subsection{Slepton mass matrix}
In the basis $\left(\tilde{e}_L, \tilde{e}_R \right)$, the slepton mass matrix is given by
\begin{align}
 m^2_{\tilde l} = 
 \begin{pmatrix}
  m^2_{\tilde l L} & \frac{v}{\sqrt 2} (T_e^\dagger \cos \beta - \mu Y_e^\dagger \sin\beta)\\
  \frac{v}{\sqrt 2} (T_e \cos \beta - \mu^* Y_e \sin\beta) & m^2_{\tilde l R}
 \end{pmatrix}\,,
\end{align}
with
\begin{align}
\notag  m^2_{\tilde l L} =& m^2_L +\frac12 v^2 \cos^2\beta Y_e^\dagger Y_e + \frac18 \Big(-(g_L^2 - g_R g_{BLR} - g_{BL} g_{RBL})v^2 \cos 2\beta \\
\notag &- (g_{BL}^2 +g_{BLR}^2 - g_R g_{BLR} - g_{BL} g_{RBL})v_R^2 \cos 2 \beta_R\Big) {\bf 1}\,,\\
\notag  m^2_{\tilde l R} =& m^2_{e^c} +\frac12 v^2 \cos^2\beta Y_e Y_e^\dagger + \frac18 \Big(- \big(g_R(g_{BLR}+g_R) + 
g_{RBL}(g_{BL}+g_{RBL}\big) v^2 \cos 2\beta  \\
&+ (g_{BL}^2 - g_R^2 +g_{BLR}^2 -g_{RBL}^2)v_R^2 \cos 2\beta_R
\Big) {\bf 1}\,.
\end{align}

\subsection{Sneutrino mass matrix}
The sneutrino mass matrix in the basis ($\tilde \nu,~\tilde \nu^c, \tilde S$) reads
\begin{align}
 M^2_{\tilde \nu} = \begin{pmatrix}
                     m^2_{\tilde \nu L} & \frac{v}{\sqrt{2}} \big( T_\nu^\dagger s_{\beta} -\mu Y_\nu^\dagger c_{\beta} \big)  
                     &  \frac{1}{2}v v_R Y_\nu^\dagger Y_S s_{\beta} s_{\beta_R}\\
                     \frac{v}{\sqrt{2}} \big( T_\nu s_{\beta} -\mu^* Y_\nu c_{\beta} \big) 
                     & m^2_{\tilde \nu R}  & \frac{v_R}{\sqrt{2}}\big(T_S s_{\beta_R} - \mu_R^* Y_S c_{\beta_R} \big) \\
                     \frac{1}{2}v v_R Y_S^\dagger Y_\nu s_{\beta} s_{\beta_R} 
                     & \frac{v_R}{\sqrt{2}}\big(T_S^\dagger s_{\beta_R} - \mu_R Y_S^\dagger c_{\beta_R} \big)
                     & ~m^2_S + \frac{v_R^2 s_{\beta_R}^2}{2} Y_S^\dagger Y_S + \frac{ (\mu_S^* + \mu_S^\dagger)(\mu_S + \mu_S^T)}{4}
                    \end{pmatrix}\,,
\end{align}
with
\begin{align}
\notag m^2_{\tilde \nu L} =& m_L^2 + \frac{v^2}{2} s_{\beta}^2 Y_\nu^\dagger Y_\nu + \\ \notag  &\frac{1}{8} 
\Big( v^2 c_{2 \beta} (g_L^2 + g_R g_{BLR}+ g_{BL} g_{RBL}) - 
 v_R^2 c_{2 \beta_R} (g_{BL}^2 + g_{BLR}^2 -g_R g_{BLR} - g_{BL} g_{RBL}) \Big) {\bf 1}\,,\\ \notag 
 m^2_{\tilde \nu R} =& m^2_{\nu^c} + \frac{v_{R}^2}{2} s^2_{\beta_R} Y_S Y_S^\dagger + \frac{v^2}{2} s_{\beta}^2 Y_\nu Y_\nu^\dagger + 
 \frac18 \Big(  v^2 c_{2 \beta} \big( (g_R-g_{BLR})g_R + g_{RBL}(g_{RBL}-g_{BL}) \big) + \\
  &v_R^2 c_{2 \beta_R}\big( (g_R-g_{BLR})^2 + (g_{BL}-g_{RBL})^2 \big)
 \Big){\bf 1}\,.
\end{align}
\end{widetext}

\end{appendix}

\end{document}